\documentclass{aa}
\usepackage[varg]{txfonts}

\usepackage{amsmath}
\usepackage{amssymb}
\usepackage{epsfig}
\usepackage{graphics}
\usepackage{color}
\usepackage{natbib}
\usepackage{caption}

\bibpunct{(}{)}{;}{a}{}{,} 

\DeclareUnicodeCharacter{00A0}{ }

\newcommand{\be}{\begin{equation}}
\newcommand{\ee}{\end{equation}}
\newcommand{\ud}{\text{d}}

\renewcommand{\vec}[1]{\ensuremath{\boldsymbol{#1}}}

\makeatletter
\def\fvec#1{\underline{\sbox\tw@{$#1$}\dp\tw@\z@\box\tw@}}
\makeatother

\usepackage[normalem]{ulem}

\newcommand{\pd}{\ensuremath{\partial}} 

\newcommand{\Req}{\ensuremath{R_{\mathrm{e}}}}
\newcommand{\sch}{Schwarzschild }
\newcommand{\Ca}{\ensuremath{\mathcal{C}}}

\newcommand{\rb}{\ensuremath{\bar{r}}}
\newcommand{\ub}{\ensuremath{\bar{u}}}
\newcommand{\wb}{\ensuremath{\bar{\omega}}}
\newcommand{\Ob}{\ensuremath{\hat{\Omega}}}
\newcommand{\nub}{\ensuremath{\bar{\nu}}}
\newcommand{\zetab}{\ensuremath{\bar{\zeta}}}
\newcommand{\Bb}{\ensuremath{\bar{B}}}
\newcommand{\mub}{\ensuremath{\bar{\mu}}}

\newcommand{\vw}{\ensuremath{v_{\omega}}}
\newcommand{\vz}{\ensuremath{v_{\mathrm{z}}}}

\newcommand{\Msun}{\ensuremath{M_{\odot}}}
\newcommand{\lgamma}{\gamma_{\text{L}}}
\newcommand{\qinv}{\ensuremath{q_{\mathrm{inv}}}}

\renewcommand{\deg}{\ensuremath{^{\circ}}}

\begin{document}

\title{Radiation from rapidly rotating oblate neutron stars}

\author{J.\,N\"attil\"a\inst{1,2} \and P. Pihajoki\inst{3} }

\institute{Tuorla Observatory, Department of Physics and Astronomy, University of Turku, V\"ais\"al\"antie 20, FI-21500 Piikki\"o, Finland \email{joonas.a.nattila@utu.fi}
  \and Nordita, KTH Royal Institute of Technology and Stockholm University, Roslagstullsbacken 23, SE-10691 Stockholm, Sweden
  \and University of Helsinki, Department of Physics, Gustaf H\"allstr\"omin katu 2a, 00560 Helsinki, Finland
}

\date{Received XXX / Accepted XXX}

\abstract{
A theoretical framework for emission originating from rapidly rotating oblate compact objects is described in detail.
Using a Hamilton-Jacobi formalism, we show that special relativistic rotational effects such as aberration of angles, Doppler boosting, and time dilatation naturally emerge from the general relativistic treatment of rotating compact objects.
We use the Butterworth--Ipser metric expanded up to the second order in rotation and hence include effects of light bending, frame-dragging, and quadrupole deviations on our geodesic calculations.
We also give detailed descriptions of the numerical algorithms used and provide an open-source implementation of the numerical framework called \textsc{bender}.
As an application, we study spectral line profiles (i.e., smearing kernels) from rapidly rotating oblate neutron stars. 
We find that in this metric description, the second-order quadrupole effects are not strong enough to produce narrow observable features in the spectral energy distribution for almost any physically realistic parameter combination, and hence, actually detecting them is unlikely.
The full width at tenth-maximum and full width at half-maximum of the rotation smearing kernels are also reported for all viewing angles.
These can then be used to quantitatively estimate the effects of rotational smearing on the observed spectra.
We also calculate accurate pulse profiles and observer skymaps of emission from hot spots on rapidly rotating accreting millisecond pulsars.
These allow us to quantify the strength of the pulse fractions one expects to observe from typical fast-spinning millisecond pulsars.
}


\titlerunning{Radiation from rapidly rotating NSs}
\maketitle

\section{Introduction}
Accurate modeling of the emission from and around compact objects is a complicated combination of radiative processes and relativity.
Not only is the object curving the space-time around it and hence affecting the trajectory of photons, but it can also affect the apparent observed radiation as the emitting surface can move with relativistic velocities.
Existing convenient and modern frameworks for the emission around rotating (typically Kerr) black holes include \textsc{geokerr} \citep{dexter2009}, \textsc{gyoto} \citep{Vincent11}, \textsc{Gray} \citep{CPO13}, \textsc{pandurata} \citep{SK13}, \textsc{astroray} \citep{SM13}, \textsc{heroic} \citep{NZP16}, \textsc{odyssey} \citep{PYY16}, and \textsc{grtrans} \citep{dexter2016}, to name a few.
Here we instead focus on the emission from rotating neutron stars, for which the Kerr metric is not a good approximation if the star is rotating rapidly.
By introducing \textsc{bender},%
\footnote{ \url{ http://github.com/natj/bender} } 
we aim to provide a similar publicly available platform for ray
tracing problems focused on spinning compact objects.

Following the path of photons in a space-time of a rotating neutron star is a challenging task, both theoretically and numerically.
Current and future observations, on the other hand, demand highly accurate models.
For example, computing accurate pulse profiles of hot spots on spinning neutron stars has recently been intensively investigated, motivated by many upcoming or planned new space-borne X-ray observatories like ESA's \textit{XIPE} \citep{XIPE},  \textit{eXTP} of
the China National Space Administration (CNSA) \citep{eXTP}, and the already deployed \textit{Astrosat} \citep{Astrosat} of
the Indian Space Research Organization (ISRO) and NASA's \textit{NICER} \citep{NICER}.
The expectation is that we may be able to better constrain the unknown neutron star (NS) equation of state (EoS) with accurate pulse profile observations, using the information encoded in the radiation \citep[see, e.g.,][]{LMB13}.

Previous studies of emission from NSs are mainly formulated in a way that uses a non-rotating curved space-time metric for the
bending of the photon paths, with special relativistic corrections to the rotational effects added separately (see, e.g., \citealt{PFC83,P95, ML98, WM01, PG03, PB06, Lamb09b, Lamb09a, LMB13, ML15}, but also \citealt{BR01} for an alternative treatment).
Space-times in these studies were also typically described by the spherically symmetric \sch metric.

As it has turned out, however, fast rotation can be a serious complication when considering the observed emission.
First of all, because a finite pressure supports the rotating star, the star is squeezed into an oblate spheroid, and the oblateness increases with increasing rotation rate \citep{CST94, MS99, MLC07, BBP13, aGM14}.
This bulge in the equator will then distort the gravitational field outside the star.
In Newtonian theory, the next-order correction to a non-spherical object (with azimuthal rotational symmetry and reflection symmetry along the equatorial plane) is defined by the quadrupole moment (see, e.g., \citealt{LP99}; but also \citealt{PA12}).
Introduction of rapid rotation will then not only make the star oblate, but will also make the exterior space-time latitude dependent.
Pioneering work in computing pulse profiles of such objects was made by \citet{CL05} and \citet{CML07}.
Recently, a general ray tracing formulation in Hartle-Thorne metric-variant was given in \citet{PJ12} and \citet{BPO12}. 
This formulation was used to compute pulse profiles in \citet{PO14}.
Here we seek to provide a similar, but open and publicly available code for solving similar types of problems.
Moreover, we focus on building a connection between the previous special relativistic formulations where different rotational effects are added separately by hand, and the full rotating general relativistic formulations where all of these effects naturally emerge from the theory.

The main focus of our framework is on the X-ray emission from accreting millisecond pulsars (AMPs) \citep{WvdK98, PW12} and nuclear-powered millisecond pulsars \citep{Watts12}.
We stress, however, that the whole framework presented in this paper is general enough to be applied to any problem of radiation originating from the vicinity of rotating compact objects. 
The radiation from AMPs emerges from hot spots on the surface of a rapidly rotating neutron star. The spots are heated by the infalling accreted material, which is being channeled to the magnetic poles by the neutron star's magnetic field.
The magnetic axis does not need to coincide with the rotational axis of the star, and hence pulsations can be observed from the spots that are rotating around the star.
In the case of nuclear-powered millisecond pulsars, quasi-coherent oscillations are observed during a thermonuclear type I X-ray burst.
The mechanism producing the pulses is, however, very similar to the case of AMPs, as an asymmetric bright patch in the burning surface layer is the origin of the observed pulsation.
Accretion can also spin up these objects into extreme rotational velocities: spin frequencies of up to 620 Hz have been verified (4U 1608$-$52; \citealt{MC02}), whereas even a typical source has a spin around $400-500$ Hz \citep{Watts12, PTR14}.
Hence, if accurate emission is to be studied from these sources, one has to take the oblate shape and (in some cases) the second-order corrections to the space-time into account.

The paper is structured as follows.
In Section~\ref{sect:theory} we introduce the framework of formulae and the theoretical background needed to compute the emission.
We also describe the numerical methods used to solve the system of equations and present the publicly available code \textsc{bender},
which implements this framework.
Next, we apply the code to various physical problems in Sect. 3.
We compare our computations with results in the literature, when possible, to verify our calculations.
Finally, in Section~\ref{sect:summary} we summarize our work.


\section{Theory}\label{sect:theory}
\subsection{Space-time metric}\label{sect:spacetime}

\begin{table*}[ht!]\label{tab:coeffs}
\begin{center}
    \caption{Series expansion terms of the metric coefficients up to $\Ob^2$.}
\begin{tabular}{l c c c c}
  \hline
  \noalign{\vskip 0.5ex}
              &  $\Ob = 0$  &  $\Ob^1$   & $\Ob^2$  &  error  \\
  \hline
  \noalign{\vskip 2ex}
  $\nub$       &  $\displaystyle \log\left[ 1-\frac{\ub}{2}\right] - \log\left[ 1+\frac{\ub}{2} \right]$ & --- & $\displaystyle +\left(\frac{\beta}{3}-qP_2(\cos\theta) \right)\ub^3 $ & $+\mathcal{O}\left(\Ob^2 \times \ub^4 \right)$ \\[3ex]
  $\Bb$         &  $\displaystyle \left( 1-\frac{\ub}{2} \right) \left(1+\frac{\ub}{2} \right)$ & --- & $\displaystyle+\beta \ub^2$ & $+\mathcal{O}(\Ob^4) \times \mathcal{O}(\ub^4)$ \\[3ex]
  $\zetab$     &  $\displaystyle \log\left[ \left( 1-\frac{\ub}{2} \right) \left(1+\frac{\ub}{2} \right) \right]$ & --- & $\displaystyle +\beta \left( \frac{3}{4}P_2(\cos{\theta}) - \frac{1}{3} \right) \ub^2$ & $+\mathcal{O}(\Ob^2) \times \mathcal{O}(\ub^4)$ \\[3ex]
  $\wb$       & --- &  $\displaystyle \wb_1 \ub^3 $ & $\displaystyle -3\wb_1 \ub^4 $ & $+ \mathcal{O}(\Ob^3) + \wb_1 \ub^3 \times \mathcal{O}(\ub^2)$ \\[2ex]
  \hline
\end{tabular}
\begin{center}{ 
    Note:
    The angular velocity term of the local inertial frame is simplified by the notation $\wb_1 \equiv 2 j/M$.
}
\end{center}
\end{center}
\end{table*}

In our following derivations we use geometric units where $G=c=1$ for the gravitational constant $G$ and the speed of light $c$.
We also assume the metric signature of $(-,+,+,+)$ following the \citet{MTW73} sign convention.
Additionally, for the actual numerical calculations in the code, we set $G M/c^2 = 1$, hence describing lengths in units of gravitational radius, where $M$ is the mass of the compact object.

The exterior space-time of a static, non-rotating, spherically symmetric mass is described by the well-known Schwarzschild metric
\be
ds^2  = -(1-u) dt^2 + (1-u)^{-1}dr^2+r^2(d\theta^2+\sin^2\theta d\phi^2),
\ee
where $r$ is the radial coordinate defined so that the area of a sphere at coordinate time $t$ is $4\pi r^2$, and we set $u \equiv 2M/r$.

This metric is equivalent to an alternative solution known as isotropic \sch metric \citep[see, e.g.,][]{MTW73}
\be
\label{eq:ISch}
ds^2 = -\left( \frac{1-\frac{\ub}{2}}{1+\frac{\ub}{2}} \right)^2 dt^2 + \left( 1+\frac{\ub}{2} \right)^4(d\rb^2 + \rb^2(d\theta^2+\sin^2\theta d\phi^2)),
\ee
where $\rb$ is the so-called isotropic radial coordinate, and we set $\ub \equiv M/\rb$.
This kind of isotropic metric has the useful feature that surfaces of constant time are conformally flat, and hence the angles are represented without distortion.
However, this also means that angular isotropic coordinates do not faithfully represent the distances within the spheres, nor does the radial coordinate correspond directly to the radial distance.
From here on, we mark all variables related to the isotropic radial coordinate with a bar on top.

We consider a rotating compact object.
To describe our system, we need a dimensionless angular velocity
\be
\Ob = \Omega \left( \frac{\Req^3}{M} \right)^{1/2},
\ee
where $\Omega$ is the angular velocity of a sphere with an equatorial radius $\Req$ and a mass $M$ scaled with the Newtonian mass shedding (Kepler) limit $(M/\Req^3)^{1/2}$ \citep[see][]{rcs}.  
Here $\Req$ is described using the usual \sch radial coordinate, and it corresponds to the equatorial radius of the star for which $2\pi\Req$ gives the proper length of the circumference in the rotational equator as measured in the local static frame.
The asymptotically flat metric near a stationary axisymmetrically rotating object in isotropic form is \citep{BW71} 
\begin{align}\begin{split} \label{eq:BWmetric}
ds^2 & = -e^{2\nub} dt^2 +
     \rb^2 \sin^2\theta \Bb^2 e^{-2\nub}(d\phi - \wb dt)^2 + \\
     & e^{2(\zetab-\nub)}(d\rb^2 + \rb^2d\theta^2),
\end{split}\end{align}
where $\wb$ is the angular velocity of the local inertial frame, and the functions $\nub$, $\Bb$ and $\zetab$ in the metric coefficients can be expanded in the powers of $\Ob$ and $\ub$ \citep{BI76}.
Here $e^{-\nub}$ is the time-dilation factor relating the proper time of the local observer to the time at infinity.
Physical interpretation of $\Bb$ follows from the fact that the proper circumference of a circle around the axis of symmetry is $2\pi(e^{-\nub} \Bb \rb \sin\theta)$.
Similarly, the interpretation of $\zetab$ follows from the fact that $e^{\zetab - \nub}$ acts as a conformal (angle preserving) factor of the space-time. 
We also note that the time and space coordinates are connected in the isotropic metric via the $\nub$-term that also enters both the radial and angular terms.
The zeroth-order terms ($\Ob = 0$) of the series expansions are the familiar \sch metric coefficients expressed in isotropic coordinates (see Table 1).

The first-order expansion in rotation ($\Ob^1$) is qualitatively related to Kerr metric.
In this case, we introduce an angular velocity term of the local inertial frame $\wb$ that accounts for the frame-dragging effects. 
It can be defined as
\be\label{eq:wbar}
\wb = \frac{2 j}{M} (\ub^3 - 3\ub^4),
\ee
where the dimensionless quantity $j=\mathcal{J}/M^2$ and $\mathcal{J} = I \Omega$ is the star's angular momentum with moment of inertia $I(M,\Omega)$.

The second-order expansion ($\Ob^2$) corresponds to a similar approximation as the Hartle-Thorne slow-rotation space-time \citep{HT68}, which introduces two quadrupole moments into the metric.  
These second-order multipole moments can be defined via the dimensionless quantities $q$ and $\beta$, the dimensionless moments of energy density and pressure, respectively.
These are, however, dependent on the selection of a coordinate system.
The coordinate invariant quadrupole moment is a combination of these two quantities and is given in \citet{PA12} (see Eq. (11) therein; see also \citealt{aGM14} and Eq. (18)) as
\be
\qinv =  q + \frac{4}{3} \beta.
\ee
This detail should be taken into account when comparing the strength of the quadrupole deviations between different metric descriptions.

\citet{YY13} showed that when the NS mass, radius, and spin are known, the star's structure (and hence the surrounding metric) is almost fully characterized by these three quantities alone. 
In other words, regardless of the unknown microphysics of the underlying matter, the NS parameters (such as moment of inertia and coordinate-invariant quadrupole moment) are connected by what is called (approximative) universal relations.
\citet{BBP13} defined empirical relations for these parameters in the Hartle-Thorne metric based on their computations of rotating NSs with various different EoS. \citet{aGM14} later refined these relations for the metric representation \eqref{eq:BWmetric} by \citet{BI76}.
In practice, we can then parameterize the previously presented quantities with great accuracy by using only the dimensionless angular velocity $\Ob$ and the compactness parameter $x$.
We note that these two parameters are defined in terms of the equatorial circumferential radius $\Req$ defined in the usual \sch coordinate system.
To the lowest order, these parameterizations are \citep{aGM14}
\be\label{eq:quad}
q = -0.11 \frac{\Ob^2}{x^2},
\ee
\be\label{eq:beta}
\beta = 0.4454 \Ob^2 x,
\ee
and
\be\label{eq:I}
I = \sqrt{x} (1.136 - 2.53 x + 5.6 x^2) M \Req^2.
\ee
We also note that both $q$ and $\beta$ are $\mathcal{O}(\Ob^2)$, whereas $j$ is $\mathcal{O}(\Ob)$.
    
It is possible to transform between the \sch coordinate radius $r$ and the isotropic $\rb$ coordinates using the relation \citep{FIP86}
\be\label{eq:rb2r}
r = \Bb e^{-\nub} \rb.
\ee

The relation between the differentials of the two radial coordinates is
\be\label{eq:drb2dr}
dr = e^{\zetab} d\rb,
\ee
which can be computed using the series representation of \cite{BI76}.

Since the series expansions of the metric coefficients are expressed in terms of the isotropic radial coordinate $\rb$, we favor this notation in our derivation.  
However, in some cases, we simplify the equations into a more intuitive form using the \sch radial $r$ coordinate.

\subsection{Oblate shape of the neutron star}\label{sect:oblate}

\begin{figure}
\centering
\includegraphics[width=9cm]{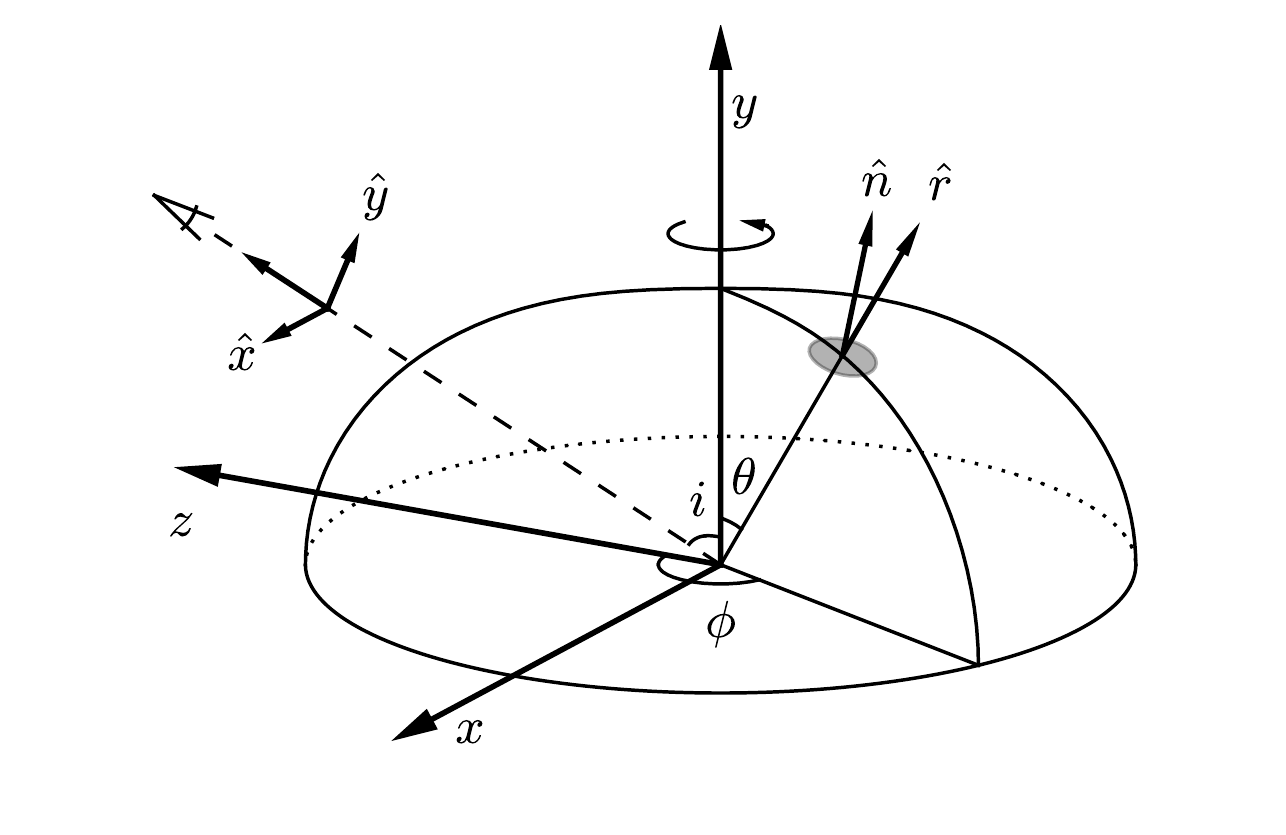}
\caption{\label{fig:geom}
Geometry of the system. Here we show the underlying spherical coordinate system with $\phi$ and $\theta$ coordinates along with the Cartesian $(x,y,z)$ coordinate system.
In addition, the observer's $\hat{x}-\hat{y}$ image plane is shown as viewed from an inclination angle $i$.
The star is taken to rotate rapidly around the $y$ -axis, which leads to an oblate (squeezed) shape for the emitting surface, and the radial vector $\vec{r}$ and the surface normal $\vec{n}$ therefore start to differ from each other.
}
\end{figure}

Because of the rotation and finite pressure supporting the NS, it is not a perfect sphere when it is rotating.  
However, it retains axisymmetry and can be approximated with an oblate spheroid.  
Similarly to the Eqs. \eqref{eq:quad}-\eqref{eq:I}, \citet{aGM14} constructed an approximate formulation for the shape of the surface of a rotating neutron star. 
It is given by expressing the radius as a function of colatitude $\theta$ as 
\begin{align}\begin{split}\label{eq:radf}
    R(\theta) &= \Req \left( 1 - \frac{\Req - R_{\mathrm{p}}}{\Req} \cos^2\theta \right) \\
              &= \Req [1-\Ob^2 (0.788 - 1.03x) \cos^2 \theta],
\end{split}\end{align}
where $R(\pi/2) = \Req$ is the radius of the star in its rotational equator, and $R_{\mathrm{p}}$ is its radius as measured along the rotation axis.

The elemental surface area for a spheroid is given as (using the usual \sch coordinates)
\be
dS(\theta) = R^2(\theta) \sin\theta \sqrt{1 + f(\theta)^2}d\theta d\phi,
\ee
where
\be
f(\theta) = \frac{1}{\bar{R}(\theta)} \frac{d \bar{R}(\theta)}{d \theta} 
= \Bb e^{-\zetab-\nub} \frac{1}{R(\theta)} \frac{dR(\theta)}{d\theta}, 
\ee
and $\Bb e^{-\zetab-\nub} \approx \left(1-\frac{2 M}{r}\right)^{-1/2} + \mathcal{O}(\Ob^2)$.
The angle $\gamma$, defined as the angle between the radial unit vector $\vec{r}$ and the surface normal $\vec{n}$ is given by
\be
\cos\gamma = \left(1 + f(\theta)^2\right)^{-1/2}.
\ee
Then the normal to the surface can be defined using the radial vector $\vec{r}$ and the tangential vector $\vec{\theta}$ as
\be\label{eq:surf_norm}
\vec{n} = \cos\gamma \vec{r} + \sin\gamma \vec{\theta}.
\ee
See Fig.~\ref{fig:geom} for a clarification of the angles.

\subsection{Geodesic motion using the Hamilton-Jacobi equation}\label{sect:hamjac}
In general, the motion of light rays in curved space-time is governed by the second-order geodesic equation.
In this section we present an equivalent theoretical formalism based on Hamilton-Jacobi (sometimes also known as super-Hamiltonian) description \citep{MTW73, cha}.
The advantage of this alternative representation is its physical intuitiveness as the formalism relies heavily on identifying and using the constants of motion of the problem.
In the end, when we apply our methods, we use both approaches (the first-order Hamilton-Jacobi equations and the second-order geodesic equations) in our calculations to show that for physically relevant problems, the results obtained are equivalent up to good numerical precision.
In a typical situation, we therefore apply the Hamilton-Jacobi method presented here, and when accuracy is the main factor, we fall back to solving the full geodesic equations.

We now discuss the motion of particles in curved space-time.
Geodesic motion in a space-time characterized by a metric $g_{ij}$ is governed by the Hamilton-Jacobi equation
\be\label{eq:hamjac}
2\frac{\pd S}{\pd \tau} = g^{ij} \frac{\pd S}{\pd x^i}\frac{\pd S}{\pd x^j},
\ee
where $g^{ij}$ is the inverse metric and $S$ denotes the Hamilton principal function.
For the two Killing vectors $t^{\alpha} = (1,0,0,0)$ (asymptotic time symmetry) and $\phi^{\alpha} = (0,0,0,1)$ (axisymmetry about the rotational axis) in rotating space-time, the Frobenius theorem implies the existence of a family of two surfaces orthogonal to these vectors \citep[see, e.g.,][]{rcs}.  
This means that there are surfaces of constant $t$ and $\phi$ in our space-time, yielding two constants of motion, namely energy $E$ and the $z$-component of the angular momentum, $L_z$.  
We then seek a solution of Eq.~\eqref{eq:hamjac} in the form
\be
S = \frac{1}{2}\delta_1 \tau - Et + L_z\phi + S_{\rb}(\rb) + S_{\theta}(\theta),
\ee
where $\delta_1$ is related to the rest-mass of the particle we study.
With the metric function \eqref{eq:BWmetric}, this becomes
\begin{align}\begin{split} 
    \delta_1 \rb^2 \Bb^2 e^{-2\nub} =~& \rb^2 \Bb^2 e^{-2\zetab} \pd_{\rb}S_{\rb}S^2 - e^{-4\nub} \Bb^2 \rb^2 (E - L_z \wb)^2 \\
                                & + \Bb^2 e^{-2\zetab} \pd_{\theta}S_{\theta}^2 + \frac{L_z^2}{\sin^2\theta}.
\end{split}\end{align}
After reorganizing terms and introducing a simplifying notation $e^{\zetab}/\Bb \equiv e^{\mub}$ , we obtain
\begin{align}\begin{split}\label{eq:S}
& e^{-2\mub}\pd_{\theta}S_{\theta}^2 + \frac{L_z^2}{\sin^2\theta} = \\ 
& \Bb^2 e^{-2\nub}\rb^2 ( e^{2(\nub-\zetab)} \pd_{\rb}S_{\rb}^2 -\delta_{1} - e^{-2\nub}(E - L_z \wb)^2 ).
\end{split}\end{align}

The individual terms in Eq. \eqref{eq:S} only depend on $r$ or only on $\theta$, that is, the dependence on $r$ and $\theta$ is separable if $\nub$, $\Bb$, and $\zetab$ only depend on $r$ and $\mu$ only depends on $\theta$. 
This is the case to first order in the stellar rotation rate $\Ob$ because $e^{\mub} = 1 + \mathcal{O}(\Ob^2)$, in addition to $\nub = \nub_0(\rb) + \mathcal{O}(\Ob^2)$ and $\Bb = \Bb_0(\rb) + \mathcal{O}(\Ob^2)$ (see Table 1). 
However, to second-order in $\Ob$, the individual terms in Eq. \eqref{eq:S} depend on both $r$ and $\theta$, that is., the dependence on $r$ and $\theta$ is not separable.
For geodesics, however, these higher-order deviations only contribute very close to the actual NS surface, and neglecting them enables us to obtain accurate approximations of the photon path.

When we now assume separability, we can introduce a separation variable $\Ca$ known as Carter's constant (third constant of motion) in order to solve the differential Eq. \eqref{eq:S}.  
By noting that the conjugate momenta correspond to the first derivatives of $S$ with respect to the generalized coordinates, we can write the components of four-momentum \fvec{p} as 
\begin{align}
  p_t        &= -E \label{eq:p_t}\\
  p_{\rb}    &= \pm e^{\zetab - 2\nub} \left( \delta_1 e^{2\nub} + (E - L_z \wb)^2 - \frac{\Ca}{\Bb^2 e^{-4\nub} \rb^2} \right)^{1/2}\label{eq:p_r}\\
  p_{\theta} &= \pm e^{\mub} \left( \Ca - \frac{L_z^2}{\sin^2\theta} \right)^{1/2}\label{eq:p_the}\\
  p_{\phi}   &= L_z\label{eq:p_p}.
\end{align}
Similarly, the components of a local tetrad frame are
\begin{align}
  p^{(t)} &= -p_{(t)} = -e_{(t)}^{\hat{\mu}} p_{\hat{\mu}} = -e^{-\nub}p_t \label{eq:tetp_t}\\
  p^{(\rb)} &= p_{(\rb)} = e_{(\rb)}^{\hat{\mu}} p_{\hat{\mu}} = e^{-\zetab + \nub} p_{\rb} \label{eq:tetp_r}\\
  p^{(\theta)} &= p_{(\theta)} = e_{(\theta)}^{\hat{\mu}} p_{\hat{\mu}} = \frac{1}{\rb} e^{-\zetab+\nub} p_{\theta} \label{eq:tetp_theta}\\
  p^{(\phi)} &= p_{(\phi)} = e_{(\phi)}^{\hat{\mu}} p_{\hat{\mu}} = \frac{1}{e^{-\nub} \Bb \rb \sin\theta} p_{\phi} \label{eq:tetp_phi},
\end{align}
where $e^{\hat{\mu}}_{(a)}$ with index $a = t, \rb, \theta$ and $\phi$ are the tetrads of metric \eqref{eq:BWmetric}.
Since we only consider null geodesics (i.e., photons), we now set $\delta_1 = 0$.

\subsection{Photon ray tracing}\label{sect:raytracing}
We now consider radiation that is emitted from the surface of the star at an emission point $(r_{\mathrm{e}},\theta_{\mathrm{e}},\phi_{\mathrm{e}})$, as seen in the static frame.
The radiation travels along a geodesic with a specific intensity $I_{E}$ as measured by an observer comoving with the emission point.  
It is observed at an image plane situated at a radial distance $r$, with $r\rightarrow\infty$.  
We then wish to calculate the projected image of the star at this image plane.

First we set up the coordinate system so that the plane of observation is toward $\phi = 0$ and $\theta = i$, where $i$ is the angle of inclination (see Fig.~\ref{fig:geom}).  
The geodesic will be emitted with a four-momentum $\fvec{p}_{\mathrm{e}}$, and if it is eventually observed at the image plane at infinity, it will have a final four-momentum of $(E,\hat{p}_r,0,0)$, purely in the radial direction.  
Likewise, the components of the position must satisfy
\begin{align}
\theta &\rightarrow i \\
\phi   &\rightarrow 0,
\end{align}
as $r\rightarrow\infty$.
The change in the time and angular components along the geodesic can be written as
\begin{align}
dt      &= \frac{p^t}{p^{\rb}}\ud \rb \label{eq:deltatime} \\
d\theta &= \frac{p^\theta}{p^{\rb}}\ud \rb \label{eq:deltatheta} \\
d\phi   &= \frac{p^\phi}{p^{\rb}}\ud \rb \label{eq:deltaphi},
\end{align}
yielding a total change of angles $\Delta\theta$ and $\Delta\phi$ when integrating from $r_{\mathrm{e}}$ to $\infty$.
The condition for being observed is then
\begin{align}
\theta_{\mathrm{e}} + \Delta\theta &= i \label{eq:thetacond}\\
\phi_{\mathrm{e}} + \Delta\phi     &= 0 \label{eq:phicond}.
\end{align}

The projected image of the star on the image plane can then be described by two celestial coordinates:
abscissa $\hat{x}$ and ordinate $\hat{y}$.
Making use of the tetrad components \eqref{eq:tetp_t}--\eqref{eq:tetp_phi}, we obtain \citep[][]{cha}
\be\label{eq:xhat}
\hat{x} = \left( \frac{rp^{(\phi)}}{p^{(t)}} \right)_{r \rightarrow \infty} = \frac{1}{\sin i} \frac{L_z}{E}
\ee
and
\be\label{eq:yhat}
\hat{y} = \left( \frac{rp^{(\theta)}}{p^{(t)}} \right)_{r \rightarrow \infty} = \frac{\sqrt{\Ca - \frac{L_z^2}{\sin^2 i}}}{E}.
\ee
Here it is useful to transform into a polar coordinate system on the image plane, as Eqs. \eqref{eq:xhat} and \eqref{eq:yhat} strongly suggest a more intuitive form if this is done. 
In this system we use as coordinates the radial distance from the center point, or the impact parameter $b$, and the polar rotation angle $\chi$.  
We take $\chi$ to increase clockwise from the projected spin axis of the neutron star, with $\chi=0$ corresponding to the projected direction from the south to the north pole of the neutron star.  
We then express the impact parameter $b$ and the angle $\chi$ via $L_z$ and $\Ca$ as%
\be
b = \frac{\sqrt{\Ca}}{E}
\ee
and
\be
\sin \chi = \frac{1}{\sin i} \frac{L_z}{\sqrt{\Ca}}.
\ee
Here, the nature of Carter's constant as a generalized squared angular momentum is apparent.
The constants of motion, combined with the geodesic null condition $p^\mu p_\mu = 0$, allow us to solve $p^\theta$ and $p^\phi$ in terms of $\rb$.
As the final step, we can substitute the four-momentum components and image-plane coordinates into Eqs. \eqref{eq:deltatime}--\eqref{eq:deltaphi} and solve the system of three first-order differential equations (in terms of $t$, $\theta,$ and $\phi$) with $\rb$ as a variable.

\subsection{Redshift and emission angle}\label{sect:redshift_angle}
It is the most convenient to define all radiative processes in the co-rotating frame of the star.
We denote variables defined in a frame that is momentarily comoving with the stellar surface with a prime.  
On the other hand, our distant observer is stationary and moving along the timelike Killing vector.  
Hence, we need to transform between stationary and rotating frames by using the four-velocity of the star's fluid.  
To make a connection to the theory of special relativity, it is convenient to define two frames: 
a corotating rest frame of the fluid $K'$ , and a non-rotating static frame $K$.
Laws of physics for the radiative transfer take the usual form in the $K'$.
A stationary observer, on the other hand, is in the non-rotating frame $K$ from where the fluid is seen to move relativistically.
We therefore need to transform between these two frames.
In addition, we need to take into account that a particle released from infinity with zero angular momentum will acquire non-zero angular velocity in the direction of the star's rotation as a
result of the dragging of inertial frames.

Four-velocity of a stationary observer with zero angular momentum (so-called ZAMO) is $o^{\alpha} = N_o (t^{\alpha} + \wb \phi^{\alpha})$, where the normalization factor $N_o = e^{-\nub}(1-\vw^2)^{-1/2}$ is obtained from $o_{\alpha}o^{\alpha} = -1$, and the velocity of the frame is 
\be
\vw = \wb \Bb e^{-2\nub} \rb \sin\theta,
\ee
indicating that the angular velocity of the ZAMO as measured by an inertial observer at infinity is $o^{\phi} / o^{t} = d\phi/dt = \wb$.

The four-velocity $s^{\alpha}$ of a circular flow can be defined using the timelike and rotational Killing vectors as $s^{\alpha} = N_s (t^{\alpha} + \Omega \phi^{\alpha})$, where the normalization factor is defined as $N_s = e^{-\nub} (1 - \vz^2)^{-1/2}$ determined by $s_{\alpha}s^{\alpha} = -1$.
Here the velocity 
\be
\vz = (\Omega - \wb) \Bb e^{-2\nub} \rb \sin\theta,
\ee
can be identified as the three-velocity measured in the frame of the ZAMO, an observer rotating with a velocity of $\vw$.

The total redshift is then given by an inner-product between a photon $u^{\alpha}$ and a four-velocity of the star's fluid $s^{\alpha}$.
With these definitions, the redshift is
\be\label{eq:redshift}
1 + z = -s_{\alpha} u^{\alpha} = e^{-\nub} ~\delta^{-1},
\ee
consisting of the gravitational part $e^{\nub}$ and of the Doppler-like factor
\be
\delta = \frac{\sqrt{1-\vz^2}}{1 - \Omega L_z}.
\ee

To compute the emission angle, we again have to take the rotating frame into account.  
This can be done by introducing a projection operator 
\be
\bar{h}_{ab} = g_{ab} + s_a s_b, 
\ee
which projects four-vectors from the non-rotating frame to the rotating frame where the radiative processes are defined.  
Our metric tensor for the rotating observer is then
\be\label{eq:proj}
\bar{h}_{ab} dx^a dx^b = e^{2(\zetab - \nub)} (d\rb^2 + \rb^2 d\theta) + \frac{\Bb^2 e^{-2\nub} \rb^2 \sin^2\theta}{1-v_Z^2} (d\phi - \Omega dt)^2.
\ee

As a definition, we can take the emission angle to be the angle between photon and a space-like surface normal vector $n_{\alpha} = N_n (0, \cos\gamma, \rb \sin\gamma, 0)$, with normalization $N_n = e^{\zetab - \nub}$ fulfilling $n_{\alpha}n^{\alpha} = -1$.  
The line element in spherical coordinates is $d\vec{s} = dr \vec{\hat{r}} + r d\theta \vec{\hat{\theta}} + r \sin\theta d\phi \vec{\hat{\phi}}$, and by combining this with Eq. \eqref{eq:surf_norm}, we obtain the presented surface normal.
When projected to the rotating frame, we can then obtain the angle from the generalized dot-product definition between two vectors as
\be\label{eq:gen_angle}
\cos\alpha' = \frac{\bar{h}_{ab}n^a u^b}{(\bar{h}_{ab} n^a n^a)^{1/2} (\bar{h}_{ab} u^a u^b)^{1/2}}.
\ee
Using the metric defined by Eq. \eqref{eq:proj} and a photon with components \eqref{eq:p_t}$-$\eqref{eq:p_p}, we obtain
\be\label{eq:cosap}
\cos\alpha' = \delta e^{2\nub-\zetab} \left[ p_{\rb} \cos\gamma + \frac{p_{\theta}}{\rb}\sin\gamma \right].
\ee
For the non-rotating observer, we similarly obtain 
\be\label{eq:cosa}
\cos\alpha = \cos\alpha' \delta^{-1},
\ee
by setting $\Omega \rightarrow 0$.
Here it suffices to notice that we are only interested in the emission angle value at the surface of the star, that is, $\rb = \bar{R}(\theta)$.
The result here is identical to the emission angle obtained with a special-relativistic approach, using flat-space trigonometry and Lorentz-boosting with $\delta$ to the rotating frame \citep[see, e.g.,][]{PB06}.

\subsection{Corotating coordinates}\label{sect:coords}
Next we define some quantities for a corotating observer located at the surface of the star.
This helps us to connect the previously presented backward-in-time method to the methods where light rays are propagated from the star to the image plane (forward-in-time methods).

Transforming from the observer's non-rotating frame $K$ to the fluid rest frame $K'$ is easily done using the previously defined projection operator $\bar{h}_{ab}$ given by Eq. \eqref{eq:proj}.
We next express this projection operator in the normal coordinate system.
Using the Eqs. \eqref{eq:rb2r} and \eqref{eq:drb2dr}, we obtain a longitudinally Lorentz-boosted metric tensor 
\be \label{eq:gammaSch} 
h_{ab} dx^a dx^b = e^{-2\nub}dr^2 + \frac{e^{2\zetab}}{\Bb^2} r^2 d\theta^2 + \lgamma^2 r^2 \sin^2\theta (d\phi - \Omega dt)^2.
\ee 
The result agrees with the \sch coordinate system up to first order in rotation because $e^{\zetab}/\Bb \approx 1 + \mathcal{O}(\Ob^2)$.  
This notation can be further simplified by defining a new azimuthal angular coordinate as $\phi' \equiv \phi - \Omega t$ that is to be used by the rotating observer.%
\footnote{We thank S. Morsink for pointing this out.}
It is important to note here that because of the rotation, the azimuthal angle $\phi$ and the time $t$ are now coupled for the comoving observer.
This new normalized longitudinal coordinate ensures that both
the rotating and the stationary observer agree that a circle drawn around the star has $2\pi$ radians.
The expression for the new longitudinal coordinate is also seen to be Lorentz-stretched by a factor of $\lgamma = (1-\vz^2)^{-1/2}$.
The corotating observer can then use this projection operator to define space-like vectors orthogonal to their world line.

Additionally, it is useful to consider another projection operator $m$ that will project from the 3D space to the 2D surface of the star.
We can define this projection as
\be\label{eq:2dmetric}
m_{ab} = g_{ab} + s_a s_b - n_a n_b = h_{ab} - n_a n_b,
\ee
where $n_a$ is the unit normal to the surface.
From here, it is easy to verify that it is perpendicular to the surface of the star as $m_{ab} n^a = 0$ and to the velocity of the surface as $m_{ab} s^a = 0$.
For simplicity, we now consider a spherical star so that the surface normal reduces to $n^a = e^{\nub} \delta_r^a$, where $\delta_a^b$ is the Kronecker delta function.
Then $m$ can be expressed as
\be
m_{ab} dx^a dx^b = \frac{e^{2\zetab}}{\Bb^2} r^2 d\theta^2 + \lgamma^2 r^2 \sin^2\theta (d\phi - \Omega dt)^2.
\ee
This is effectively a 2D metric tensor of the star's surface that is perpendicular to the world line of the corotating observer.

Definitions here are also relevant for the so-called Schwarzschild+Doppler (S+D) approximation \cite[see, e.g.,][]{PB06}.
In the S+D approximation, the observer's polar coordinate plane $(b,\chi)$ is connected to the corotating spherical coordinates $(\phi', \theta')$ of the star.
We note that this connection is done algebraically in the S+D method and so there is no need to determine the full path of the ray using partial differential equations as the problem reduces to calculating the so-called lensing integral alone.
The S+D calculations are done in a special relativistic framework where quantities are defined in a corotating coordinate system that are then Lorentz-boosted into the static non-rotating frame.
We now show the correct expression of this change of frame so
that the results match the backward-in-time method (see also \citealt{CML07}, for an alternative derivation).
In the usual \sch metric, the impact parameter $b$ can be obtained as a function of the emission angle $\alpha$ given by Eq. \eqref{eq:cosa} as
\be
b = \frac{1}{\sqrt{1-u}} R \sin\alpha\ee
by setting $\cos\gamma \rightarrow 1$, $\sin\gamma \rightarrow 0$ (spherical star), and $\omega \rightarrow 0$ along with $e^{-2\nu} \rightarrow (1-u)^{-1/2}$ (\sch metric, i.e., $\mathcal{O}(\Ob^1)$).
Here we use the emission angle as measured by the non-rotating observer in frame $K$.
In this case, the solid angle is then simplified to 
\be
d\Omega_o = bdb \, d\chi = \frac{1}{1-u} \frac{d \cos\alpha}{d \cos\psi} R^2 \cos\alpha ~ \frac{d\chi d\cos\psi}{D^2},
\ee
where $\psi$ is the lensing angle.
Using the spherical symmetry of the \sch metric, we can directly connect the $\psi$ and $\theta$ along with $\chi$ and $\phi$ to obtain
\be
d\Omega_o = \frac{1}{1-u} \frac{d \cos\alpha}{d \cos\psi} ~\cos\alpha ~ \frac{dS}{D^2},
\ee
where $dS = R^2 \sin\theta d\theta d\phi$ is the area element for the non-rotating static observer in $K$.
Using $d\Omega_o$ to compute the flux, we would then obtain the observed (received) flux valid for the observer in frame $K$.

Integration over some finite-sized features, on the other hand, is done on the surface of the star in the corotating frame $K'$.
This results in a mixing of the different frames because we would like the integration to happen simultaneously for the observer in $K$, that is to say, we need a connection between $t$, $\phi,$ and $\phi'$.
To do this, we need to define the differential area element in the corotating frame.
This can be obtained by considering the 2D metric tensor $m$ of the surface given by Eq. \eqref{eq:2dmetric}.
Using this, we can derive the corotating differential area element as
\be\label{eq:dSp}
dS' = \sqrt{\det m_{ab}} dx^a dx^b = \frac{e^{\zetab}}{\Bb} \lgamma R^2 \sin\theta d\theta (d\phi - \Omega dt).
\ee
This means that the rotation will result in a stretching of the area element by a factor of $\lgamma$ , whereas the quadrupole moments will deform it by a factor of $e^{\zetab}/B$.
This result is obtained purely from differential geometry.

Next we work only in the \sch metric to draw a direct connection to the S+D approximation.
Using Eq. \eqref{eq:dSp}, we obtain\be\label{eq:dS_subs}
dS' = \lgamma R^2 \sin\theta d\theta (d\phi - \Omega dt) = \lgamma R^2 \sin\theta' d\theta' d\phi',
\ee
as given in the corotating coordinates defined as $\theta' \equiv \theta$ and $\phi' \equiv \phi - \Omega t$.
This differs by a factor of $\lgamma$ from an incorrect result that would be obtained by erroneously assuming that $dS' = R^2 \sin\theta' d\theta' d\phi'$.
To transform from the non-rotating $K$ frame to the corotating frame $K'$ , we can use the Lorentz invariance of the photon beam cross-section given as \citep{Terrell60, LB85}
\be
\cos\alpha ~dS = \cos\alpha' ~dS'.
\ee
The connection between the emission angles $\cos\alpha'$ and $\cos\alpha$ is also known from Eqs. \eqref{eq:cosap} and \eqref{eq:cosa} and is seen to be a simple Doppler boost factor $\delta$.
We then obtain
\be\label{eq:dSisdSp}
dS = \delta dS'.
\ee
Finally, the total observed angular size is then seen to be 
\be
d\Omega_o = \frac{\lgamma \delta}{1-u} \frac{d \cos\alpha}{d \cos\psi} ~\cos\alpha ~ \frac{R^2 \sin\theta' d\theta' d\phi'}{D^2}.
\ee

\subsection{Emission}\label{sect:emission}
The observed (i.e., received) flux at photon energy $E$ from a small area on an image plane is
\be
dF_E = I_E d\Omega_o,
\ee
where $I_E$ is the specific intensity of the radiation at infinity, and $d\Omega_o$ is the solid angle subtended by the element as measured by the observer. 
The total flux is then the integral of these elements over the image plane.
As a final step, this observed flux has to be connected to the actual emerging radiation.

From Eq.~\eqref{eq:redshift}, the relation between the emergent energy $E'$ to the observed energy $E$ is $E/E' = (1 + z)^{-1}$.
The connection between the monochromatic observed and local intensity is then (see, e.g., \citealt{MTW73}; \citealt{RL79})
\be
I_E = \left( \frac{E}{E'} \right)^3 ~I'_{E'}(\alpha'),
\ee
where $I'_{E'}(\alpha')$ is the intensity computed in the frame comoving with the emitting area.
The radiation here is emitted in the direction of the angle $\alpha'$ defined in the local rotating frame.
Integrating over the energies, we obtain the bolometric intensity
\be
I = \left(\frac{E}{E'} \right)^4 ~I'(\alpha').
\ee
The total (monochromatic) flux as a function of the observer's time $F_E(t)$ can then be obtained by integrating over the whole image,
\be\label{eq:fluxint}
F_E(t) = \int I_{E}(t) ~d\Omega_o = \int\int \frac{I'_{E'}(t_*, \alpha')}{(1+z)^3}  ~\frac{bdb \, d\chi}{D^2}
,\ee
where $t_* = t - \Delta t$ is the time when the photon was emitted as measured in the non-rotating coordinate system.
This can be computed when we know the total travel time $\Delta t$ against some reference photon, for example, the one with the shortest path to the observer.

All of these quantities on the (non-rotating) spherical coordinate system $(\theta, \phi)$ are then mapped to the observer's polar image coordinates $(b, \chi)$ via ray tracing. The original longitudinal coordinate of the emission is easily obtained from $\phi_{\mathrm{e}} = \phi - t_* \Omega$ because both $t_*$ and $\Omega$ are defined for a distant observer, and change in the azimuthal coordinate is Lorentz invariant.
This allows us to connect the observables that our distant observer will see to the local rest frame of the gas where most of the physical processes are naturally defined.

\subsection{Angular distribution of radiation}\label{sect:angular_distr}
\subsubsection{Blackbody radiation}

For pulse profile calculations, the simplest angular distribution of radiation is the isotropic radiation.
Here we consider blackbody emission described by the specific intensity
\begin{equation}
  B_{\mathrm{E}}(T) = 5.04 \times 10^{22} \frac{E^3}{\mathrm{exp}(E/T) -1} \,\mathrm{erg}\,\mathrm{s}^{-1}\,\mathrm{cm}^{-2}\,\mathrm{keV}^{-1}\,\mathrm{sr}^{-1},
\end{equation}
where $T$ and $E$ are given in keV.

\subsubsection{Atmosphere dominated by electron scattering}

Next we consider beamed radiation.
For simplicity, we continue to assume that the spectral distribution is given by the Planck function, but now we assume that the angular distribution corresponds to that given by coherent electron scattering in a plane-parallel, semi-infinite (optical depth $\tau \rightarrow \infty$) atmosphere.
This beaming pattern is described by the so-called Hopf function $H(\mu)$.
Introducing a variable $\mu \equiv \cos\alpha'$, the result is
\begin{equation}\label{eq:hopf}
  I_{\mathrm{E}}(\mu) = B_{\mathrm{E}}(T) \frac{H(\mu)}{2\alpha_1},
\end{equation}
where
\begin{equation}
  \alpha_n = \int_0^1 H(\mu) \mu^n d\mu
\end{equation}
are the moments of the function $H(\mu)$, which is a solution of the Ambartsumian-Chandrasekhar integral equation \citep[see, e.g.][]{Cha60,Sob63}
\begin{equation}\label{eq:hmu}
  H(\mu) = 1 + \mu H(\mu) \int_0^1 \frac{\Psi(\eta)}{\mu + \eta} H(\eta) d\eta.
\end{equation}
Here $\Psi(\mu)$ is the characteristic function, which depends on the scattering law considered.
For Rayleigh (Thomson) scattering, it is
\begin{equation}
  \Psi(\mu) = \frac{3}{16}(3-\mu^2).
\end{equation}
Given $\Psi$, the integral equation \eqref{eq:hmu} can then be iteratively solved, for example, by computing
\begin{equation}\begin{split}
    H_{n+1}(\mu) =  \frac{1}{2} H_n(\mu) + \frac{1}{2}\Biggl(& \sqrt{1-2\int_0^1 \Psi(\eta)d\eta} \\
                    &+ \int_0^1 \frac{\eta \Psi(\eta)}{\mu + \eta} H_n(\eta) d\eta \Biggr)^{-1},
\end{split}\end{equation}
with a starting guess of
\begin{equation}\label{eq:apprx_hopf}
  H_0(\mu) = 1 + 2.3\mu - 0.3\mu^2.
\end{equation}

We note that Eq. \eqref{eq:hopf} is physically inconsistent (blackbody radiation must be isotropic), but we adopt this spectrum and beaming pattern for the sake of illustration. 
Electron-scattering atmospheres can produce spectra that have spectral shapes similar to a Planck function, but they are much less efficient. 
Eq. \eqref{eq:hopf} can describe such emission approximately, but only if it is preceded by an efficiency factor that depends on the color-correction factor $f_{\mathrm{c}}$ as $f_{\mathrm{c}}^{-4} \approx 0.15$ \citep[see, e.g.,][]{SPW11, SPW12}.
We also note that even the simple polynomial expansion \eqref{eq:apprx_hopf} has an accuracy better than $<2\%$, and it can therefore be used in approximate solutions with a corresponding first moment of $\alpha_1 = 1.19167$.

\subsection{Method of solution}\label{sect:num_methods}

\begin{figure}
\centering
\includegraphics[width=8cm]{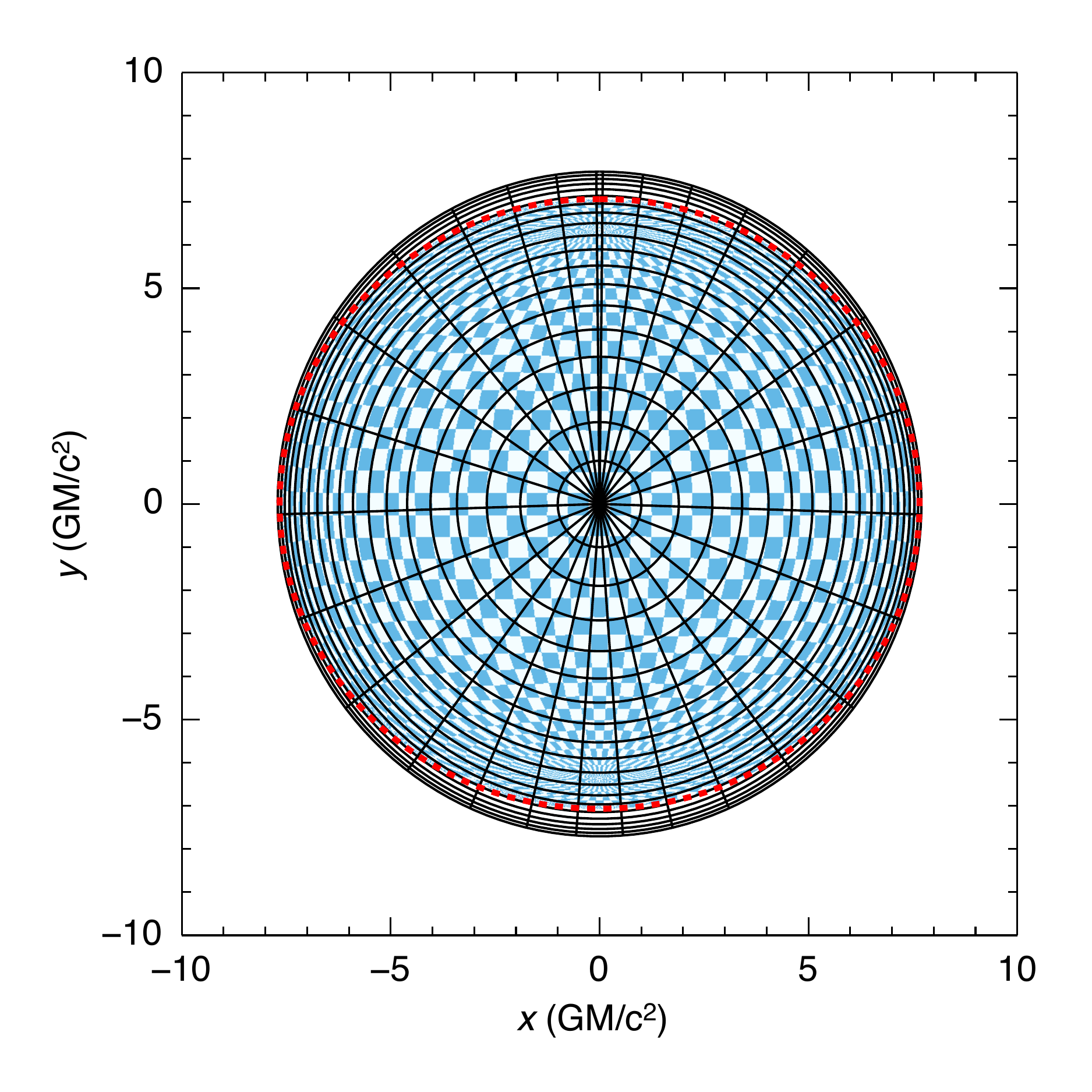}
\caption{\label{fig:grid}
Example of a non-equidistant polar grid used in our ray tracing with $N_r = 20$ and $N_{\chi} = 30$ points.
    The red dashed line corresponds to the outline of the actual oblate star that is covered with a chessboard pattern.
}
\end{figure}

In practice, when calculating the observed time-dependent emission, we have to 
\begin{enumerate}
    \item set up the image plane, 
    \item propagate the geodesics using either the full equations of motion or the approximate Hamiltonian-Jacobi result, and
    \item compute the actual number of photons received now that we know the connection between $(b,\chi$) and $(\theta, \phi_e)$.
\end{enumerate}

We trace the photons from the image plate at infinity all the way to the surface of the star by solving the first-order differential Eqs. \eqref{eq:deltatime}--\eqref{eq:deltaphi} or the full geodesic equations backward in time.  
In our numerical computations, we place the image plane at $\sim 10^5 R_{\mathrm{e}}$.
For the Hamilton-Jacobi formalism, we make a simple variable substitution $\tilde{x} = 1/r$
 that helps us by stretching the step when
far from the star and shortening it when approaching the star
surface.  Here an adaptive step size second-order Heun Runge-Kutta
integrator is used with the forward Euler method as
the predictor and trapezoidal method as the corrector.  
All photons that travel more than $1.05 R_{\mathrm{e}}$ away from the star after a U-turn are
terminated and considered to have missed the star.
The full geodesic equations are solved using the \textsc{arcmancer} code \citep[see][and the related equations therein]{PRJ16}.

Our image plate is defined using a polar coordinate system with a radial
coordinate $b$ (i.e., the impact parameter) and an angular coordinate $\chi$.
We also employ a non-equidistant grid in both coordinates to accommodate
the extra resolution needed around the edges of the star.  The
radial
coordinate $b$ is defined using a Gauss-Laguerre abscissa (i.e.,
$e^{-b}$-weighted), and the angle coordinate is weighted with a simple
sinusoidal function so that the resolution is increased around the top
and bottom parts where $\chi = 0$ or $\pi$, which is near the location of the poles
(see Fig.~\ref{fig:grid}).  By ray tracing, we then obtain
a mapping between the image plane and the surface of the star, defined
on a grid.
Arbitrary positions in the image plane are obtained by a quadratic interpolation in $(b, \chi)$ space.

For pulse profile calculations where only a small part of the star is
emitting, we first search a crude location of the spot on the image plane
and then impose a fine subgrid around it in order to accurately
calculate the flux from this small patch.  The subgrid itself is defined
either in a polar grid (by constraining minimum and maximum $\chi$ and
$b$) or in a Cartesian image grid (by constraining minimum and maximum
$\hat{x}$ and $\hat{y}$) depending on the total area covered in the observer's sky.  To
calculate the total flux $F(t),$ we then integrate this small subgrid by
using an adaptive multidimensional integration.  
The algorithm is based on a tensor product of nested Clenshaw-Curtis quadrature rules and is implemented using the \textsc{Cubature} package.\footnote{
\url{http://ab-initio.mit.edu/wiki/index.php/Cubature}}

Such a general way to treat the problem of course also has its disadvantages.
Ray tracing photons in general is a computationally very expensive problem.
For fast calculations, other more approximate ways exist to solve the problem, such as the oblate \sch method, where the symmetries of the \sch space-times are extensively used and the ray tracing reduces to lensing angle integrals \citep[see, e.g.,][]{PB06, MLC07}.
We emphasize that our focus is not to compete with these methods in speed, but to verify their results using a more general description of the problem.

\begin{figure*}
\centering
\includegraphics[clip, trim=0.0cm 0.5cm 0.0cm 2.0cm, width=9cm]{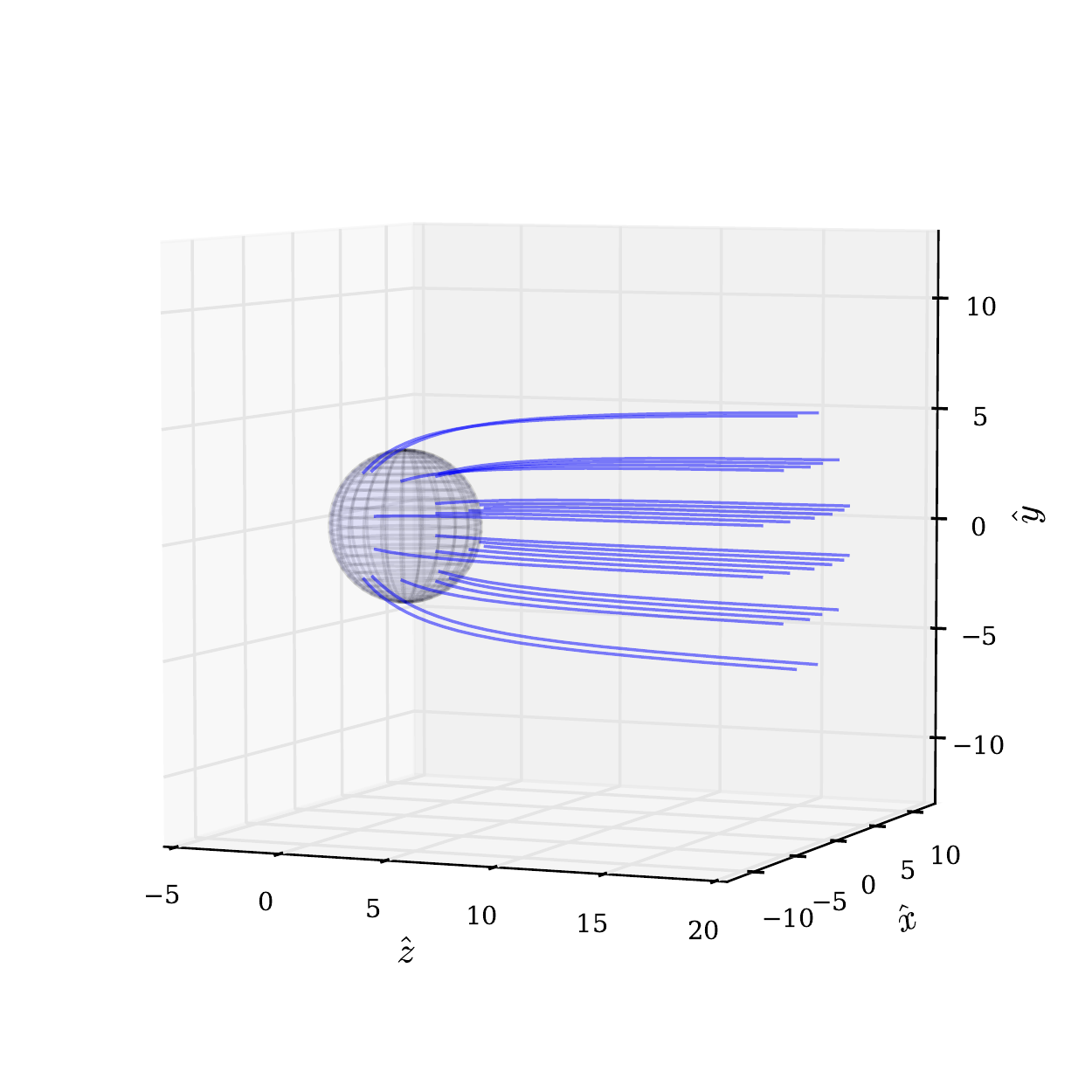}
\includegraphics[width=8cm]{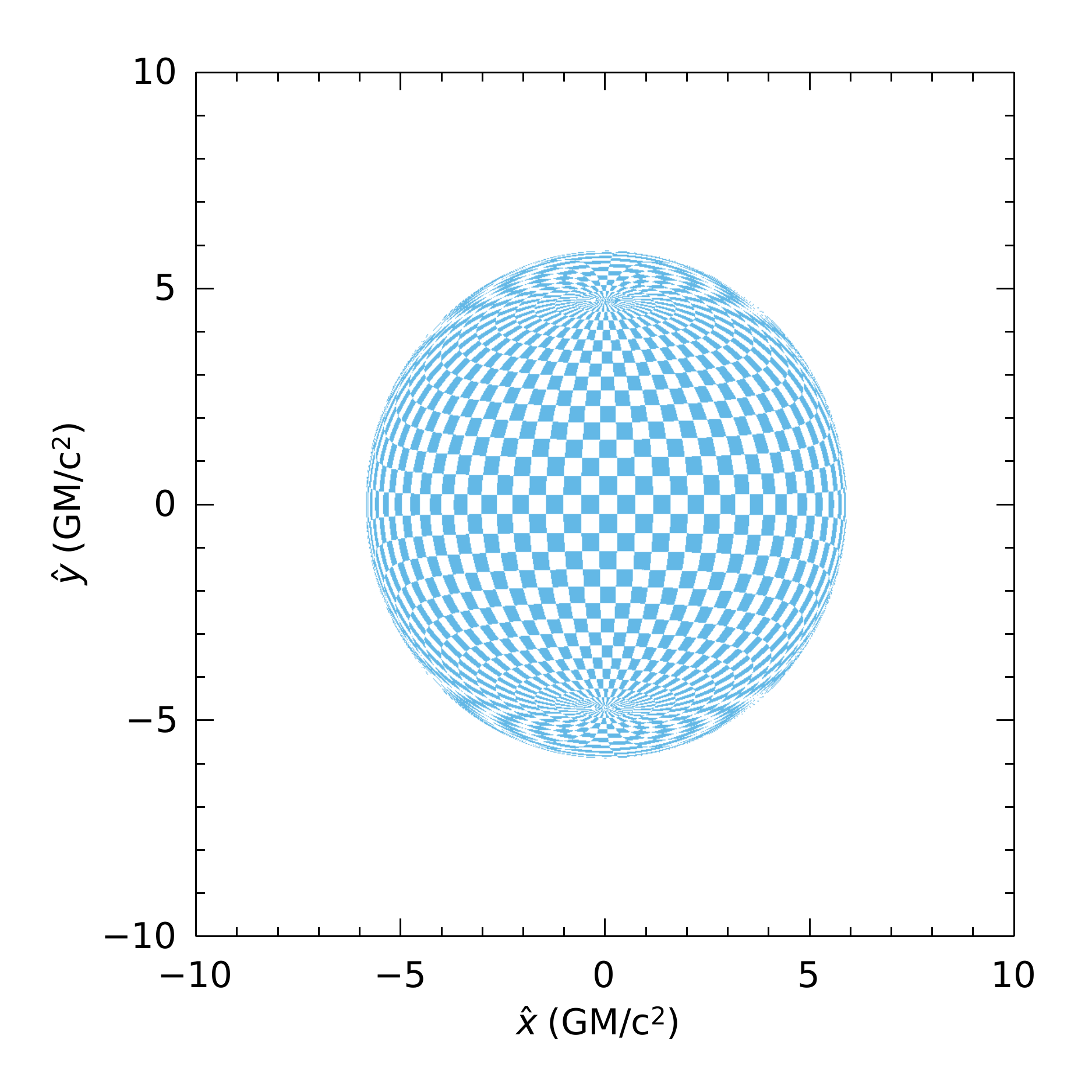}
\caption{\label{fig:image}
  Formation of an image in curved space-time.
  The left panel shows a 3D visualization of the photon trajectories in the curved space-time starting from the star and ending at the observer's image plane.
  The right panel shows the image that the observer sees using the Cartesian $\hat{x}$ and $\hat{y}$ coordinates.
  For illustrative purposes, the neutron star surface is covered with a chessboard pattern.
  }
\end{figure*}

\section{Applications and verification}\label{sect:appl}

Next we present some applications of the framework to some simple physical problems related to neutron stars to showcase possible applications of the code.
The examples are also meant to act as a further verification, as we provide a comparison with existing literature results, when possible.
Most notably, we perform an extensive comparison against AMP pulse profiles computed with a forward-in-time method as presented in \citet{PB06} and \citet{aGM14}.
Our code serves as a great cross-verification tool for these types of special relativistic formulations because our framework is fully general relativistic and propagates photons backward in time from the observer to the surface.

\subsection{Images of neutron stars}

As a first application of the code, we can determine the photon trajectories using the ray tracing algorithm and produce an image of the neutron star as as seen by the observer.
This also shows how we can connect the Cartesian coordinates $\hat{x}$ and $\hat{y}$ of the observer to the coordinates $\phi_{\mathrm{e}}$ and $\theta_{\mathrm{e}}$ of the star.
The left panel in Fig.~\ref{fig:image} shows the trajectory of the photons in 3D, using a $\hat{z}$-coordinate in addition to the Cartesian image plane coordinates $\hat{x}$ and $\hat{y}$.
The star is chosen to have $R=12\,\mathrm{km}$, $M=1.5\,\Msun$, and has a spherical shape, whereas the observer is located at the equator with an inclination of $i = 90^{\circ}$.
Here the photons originate from the image plane located at $\hat{z} = 20$ and are then propagated backward in time until they intersect with the surface of the star (center of the star located at $\hat{z} = 0$) , visualized with a spherical see-through wire-grid frame.
The right panel of the figure shows the projected image as seen by a distant observer.
Here the star is covered with a chessboard pattern to show how the $\phi_{\mathrm{e}}$ and $\theta_{\mathrm{e}}$ coordinates on the neutron star surface are seen by the distant observer.
In case of no rotation ($f = 0\,\mathrm{Hz}$), the image outline is verified to be mirror symmetric with respect to reflection along the $\hat{x} = 0$ vertical axis and along the $\hat{y} = 0$ horizontal axis of the image.

\subsection{Accuracy of the split Hamilton-Jacobi propagator}

\begin{figure*}[htbp!]
\centering
\includegraphics[width=18cm]{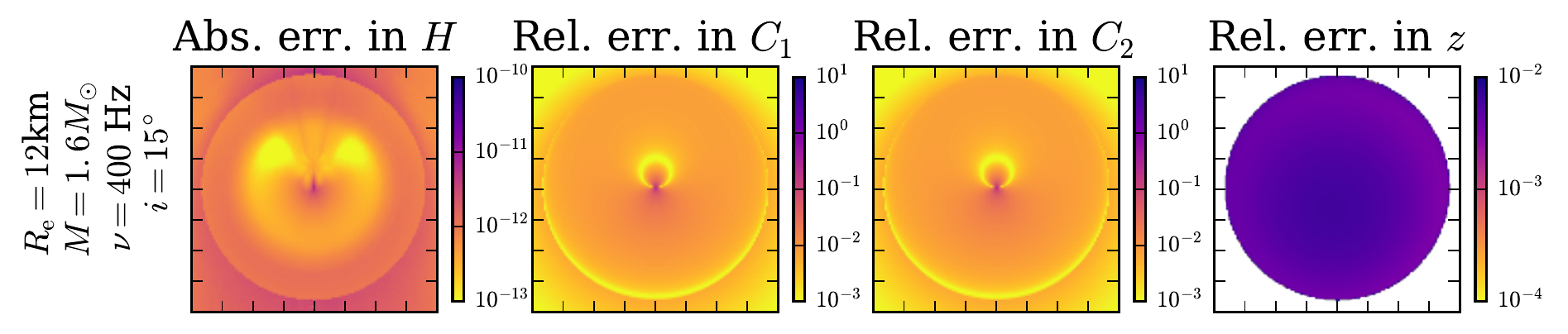}
\includegraphics[width=18cm]{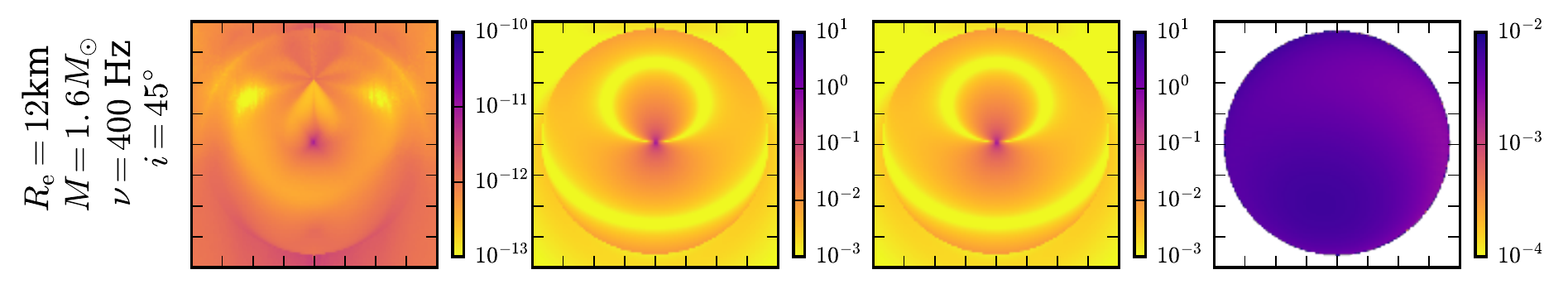}
\includegraphics[width=18cm]{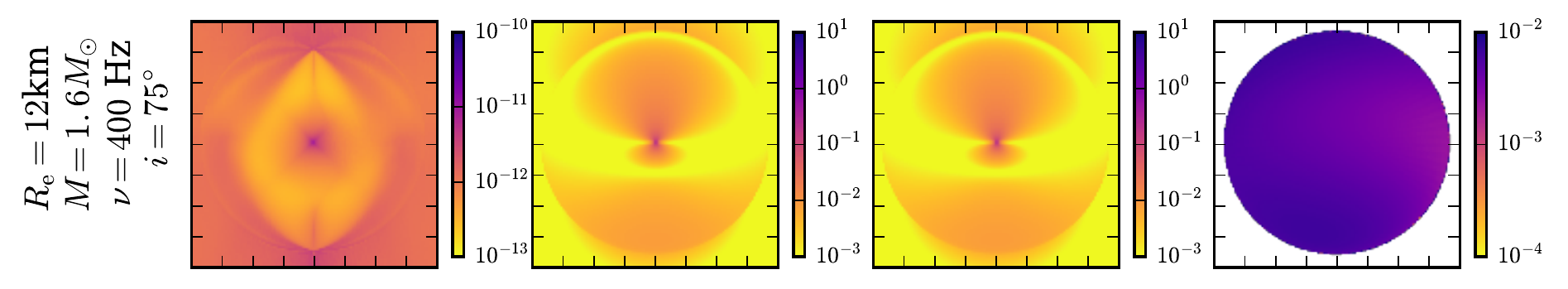}
\includegraphics[width=18cm]{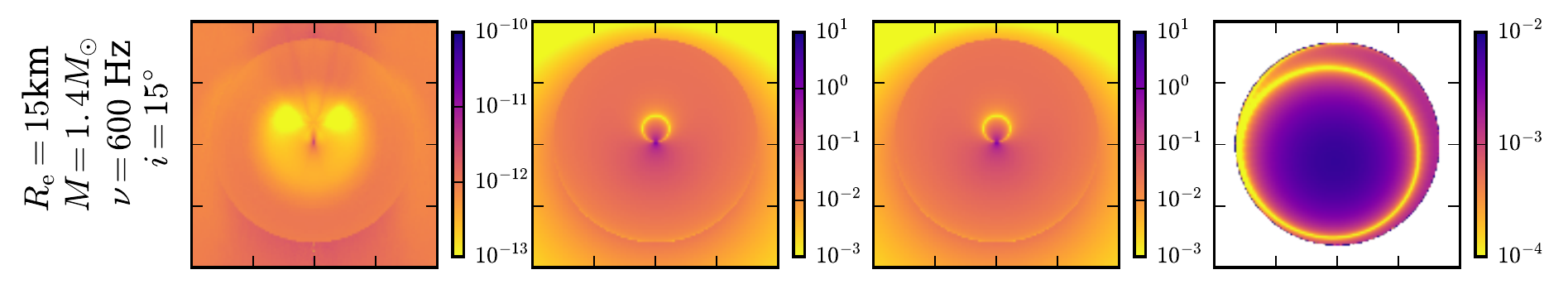}
\includegraphics[width=18cm]{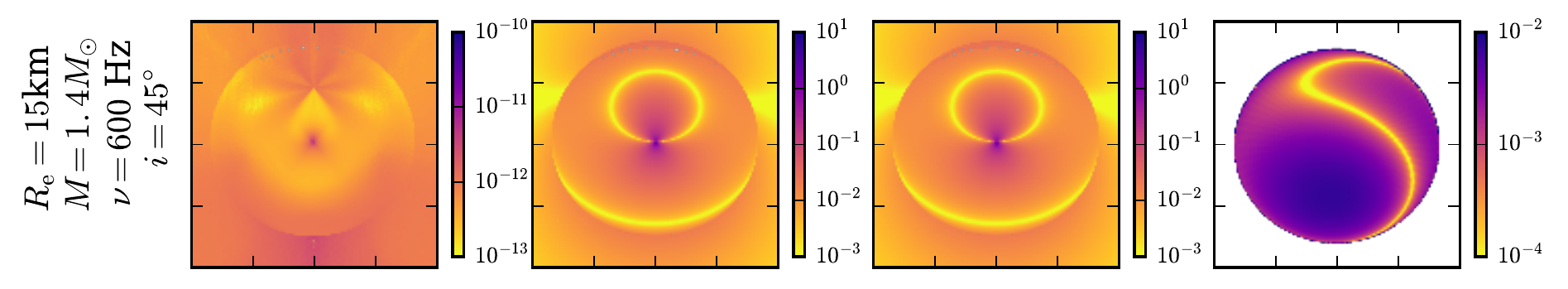}
\includegraphics[width=18cm]{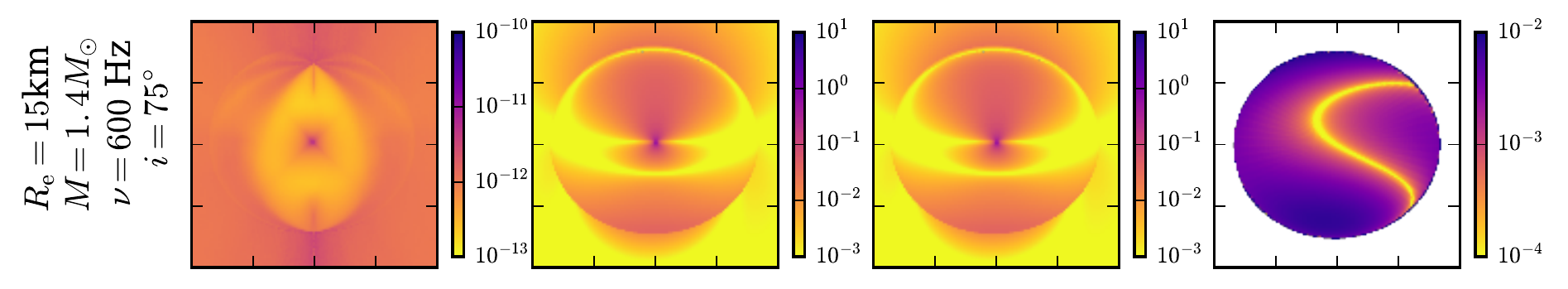}
\caption{\label{fig:H_C1_C2}
    Errors in ray tracing two neutron stars with $R_{\mathrm{e}}=12\,\mathrm{km}$, $M=1.6\,\Msun$, $\nu=400\,\mathrm{Hz}$ or $R_{\mathrm{e}}=15\,\mathrm{km}$, $M=1.4\,\Msun$, $\nu=600\,\mathrm{Hz}$, and with observer inclinations $i=15^\circ$, $45^\circ$ and $75^\circ$, solving the full geodesic equation vs. the split Hamilton-Jacobi equation.
    The leftmost panels show the maximum variation in the Hamiltonian $H$ of the geodesic, while the two center panels show the maximum variation in the Jacobi constant, computed either with the left side ($C_1$) or the right side ($C_2$) of Eq. \eqref{eq:S}.
    The rightmost panels show the relative error in the redshift computed by solving the split Hamilton-Jacobi equation compared with the redshift computed by solving the full geodesic equation.
}
\end{figure*}

Next, we study the feasibility of the split Hamilton-Jacobi propagator by comparing to results from the general-purpose geodesic solver \textsc{arcmancer} \citep{PRJ16}.
The comparison solver directly solves the Lagrangian equations of motion of the geodesic in an arbitrary user-given metric. 
We compute the change in the Jacobi constant as computed with both sides of Eq. \eqref{eq:S}, and the change in the value of the Hamiltonian of the geodesic for two different neutron star configurations:
$R_{\mathrm{e}}=12\,\mathrm{km}$, $M=1.6\,\Msun$, and $\nu=400\,\mathrm{Hz}$, and $R_{\mathrm{e}}=15\,\mathrm{km}$, $M=1.4\,\Msun$, and $\nu=600\,\mathrm{Hz}$.
The observer inclinations are $i=15^\circ$, $45^\circ$ , and $75^\circ$.
The error in the Hamiltonian or in the Jacobi constant reflects the error in the photon path.
To study the effect this has on the actual observables, we compare the values of the photon redshifts $z$ obtained either with the full numerical propagator or with the split Hamilton-Jacobi propagator.
Deviations in this value as a function of the location then reflect the error not only in $z$ itself, but also in the $\phi_e$ and $\theta_e$ coordinates.
The results are shown in Fig.~\ref{fig:H_C1_C2}. 

In general, we see that the assumption of separability, as measured by the variation in the Jacobi constant, is good to a level of $10^{-3}$--$10^{-2}$, except for geodesics that hit the center of the neutron star from the observer's point of view. 
However, the examples were deliberately chosen to be extreme, and the approximation of separability is much better for more slowly rotating neutron stars.
Furthermore, the relative error in the observed redshift is always smaller than $7 \times 10^{-3}$, even for these extreme cases.

This small error has two reasons.
First, the splitting of the Hamilton-Jacobi equation is an excellent approximation because the quadrupole moment produces a deviation in the metric only very close to the star.
When photons are propagated from a distant location to the stellar surface, the effect on the trajectory is negligible.
Second, the splitting does not affect the redshift calculations as we use the exact form of $1+z$ as given by Eq. \eqref{eq:redshift}.
When these two aspects of the method are combined, we observe the excellent agreement against the results obtained from the full geodesic equations.
Last, we note that in cases of extreme rotation or when high precision is required, the geodesic propagation can easily be made using the \textsc{arcmancer} instead of Eqs. \eqref{eq:deltatime}--\eqref{eq:deltaphi}, while using the rest of the results in this paper for the actual radiation computations.

\subsection{Line profiles}

\begin{figure}
\includegraphics[width=8cm]{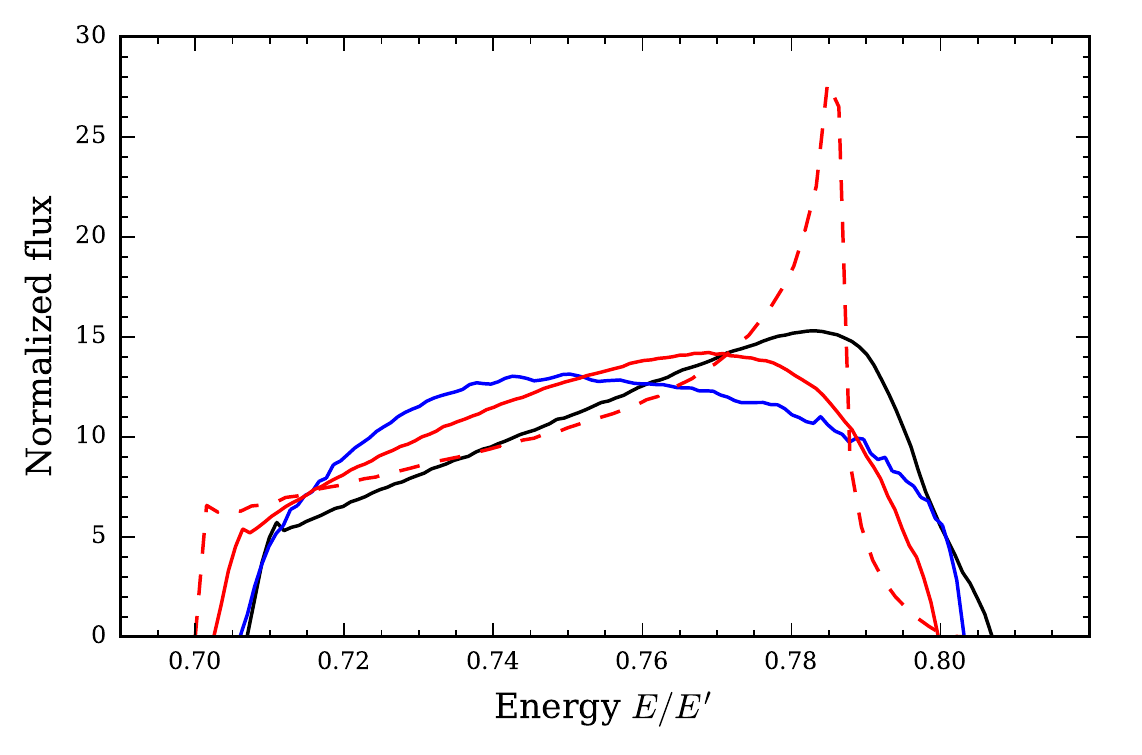}
\caption{\label{fig:line_profiles}
Line profiles from a star with $R = 10\,\mathrm{km}$, $M = 1.4\,\Msun$, and a rotation frequency of $700$ Hz seen by an observer at an inclination $i = 20^{\circ}$ computed by solving the geodesic equations using \textsc{arcmancer}.
The black line shows the profile of a spherical star with a Schwarzschild exterior space-time. 
The blue line represents the profile of an oblate star with $R_{\mathrm{e}} = 10\,\mathrm{km}$ and a Schwarzschild exterior space-time. 
The red solid line denotes the profile of an oblate star with an exterior space-time that has a non-zero quadrupole moment (see text). 
    The red dashed line shows the profile of an oblate star with a quadrupole moment that has been artificially increased by a factor of 4 (see text).
}
\end{figure}


\begin{figure*}[htbp!]
\centering
    \includegraphics[width=18cm]{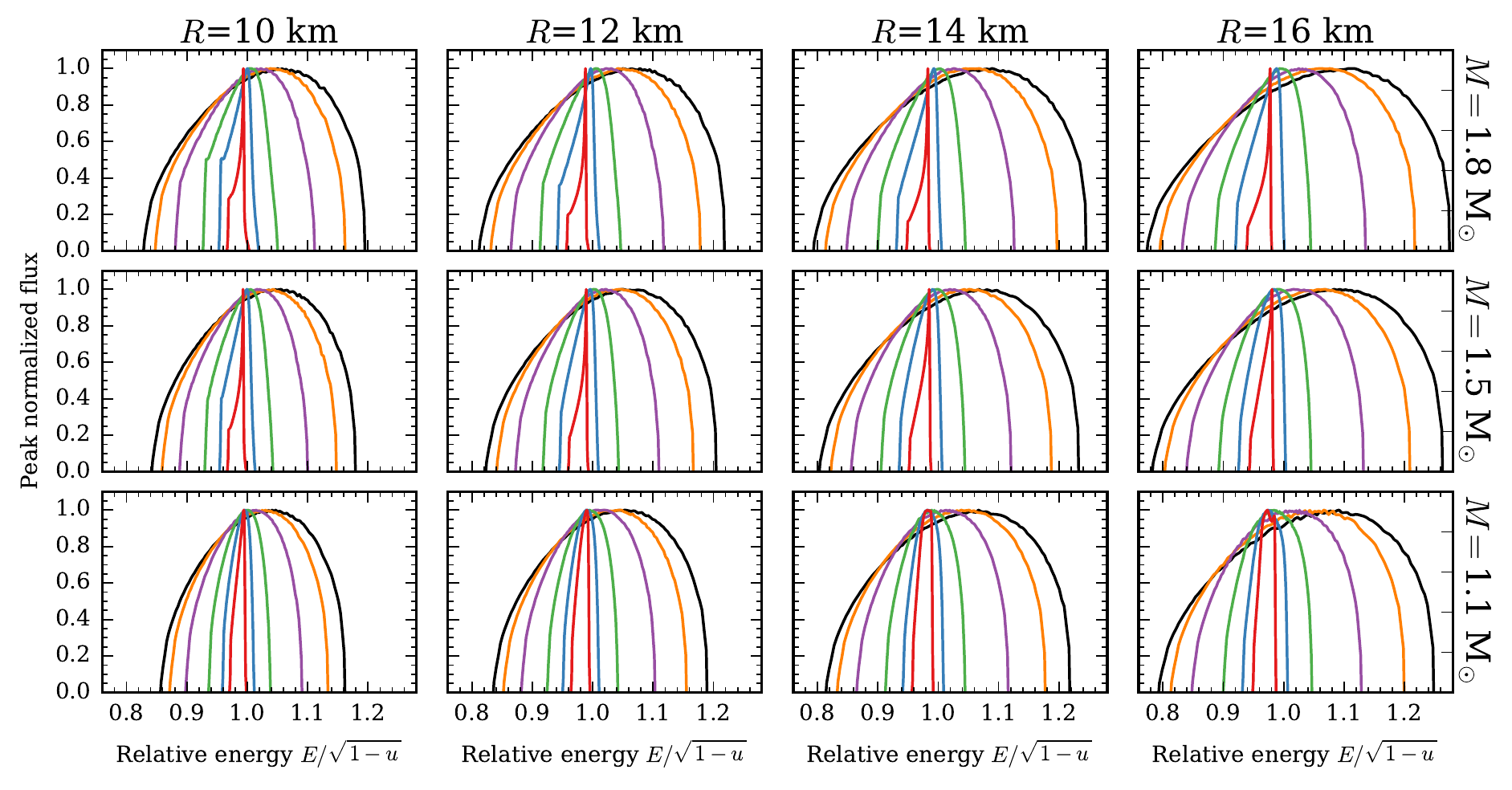}
\caption{\label{fig:sweep}
    Exact shapes of line profiles for different neutron stars spinning at $600\,\mathrm{Hz}$.
Observer inclinations span a range from $i=5\deg$ (red), $10\deg$ (blue), $20\deg$ (green), $40\deg$ (purple), and $60\deg$ (orange), to $90\deg$ (black).
The energy in the horizontal axis is scaled with the compactness $\sqrt{1-u} = \sqrt{1-2GM/R_{\mathrm{e}}}$ that would be expected from the gravitational redshift alone.
Likewise, the flux in the vertical axis is always normalized with the peak flux to better show the evolution of the profile shape.
}
\end{figure*}

\begin{figure*}[htbp!]
\centering
    \includegraphics[width=18cm]{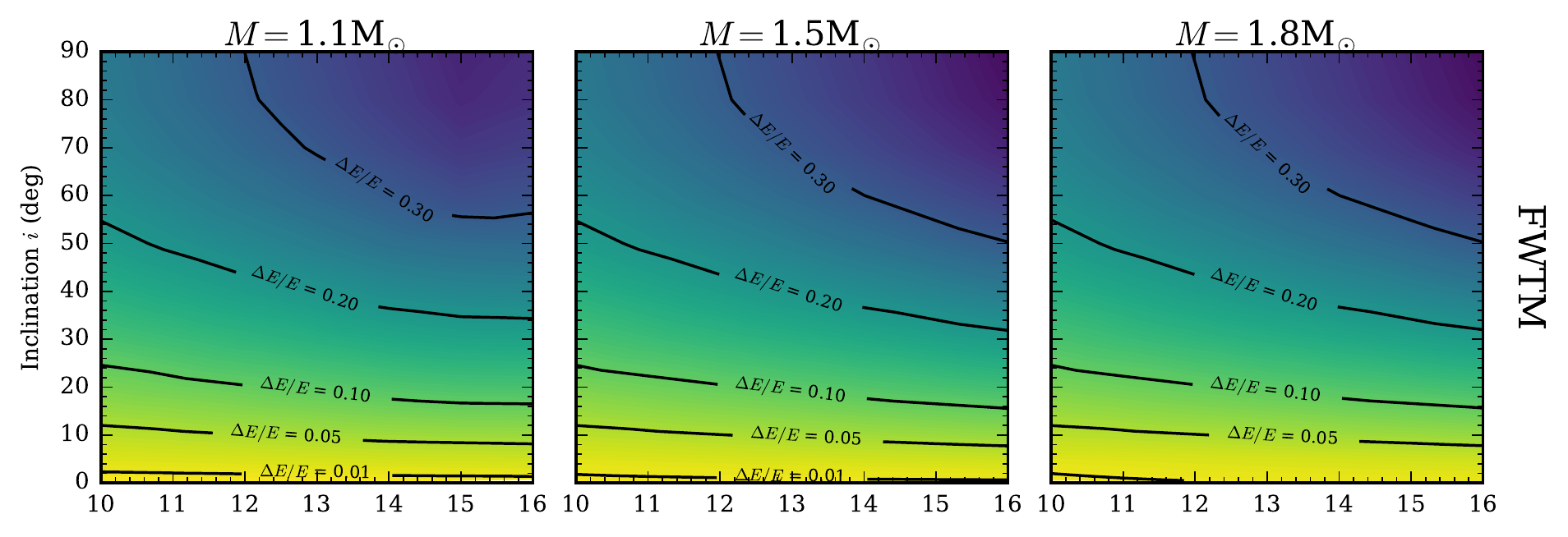}
    \includegraphics[width=18cm]{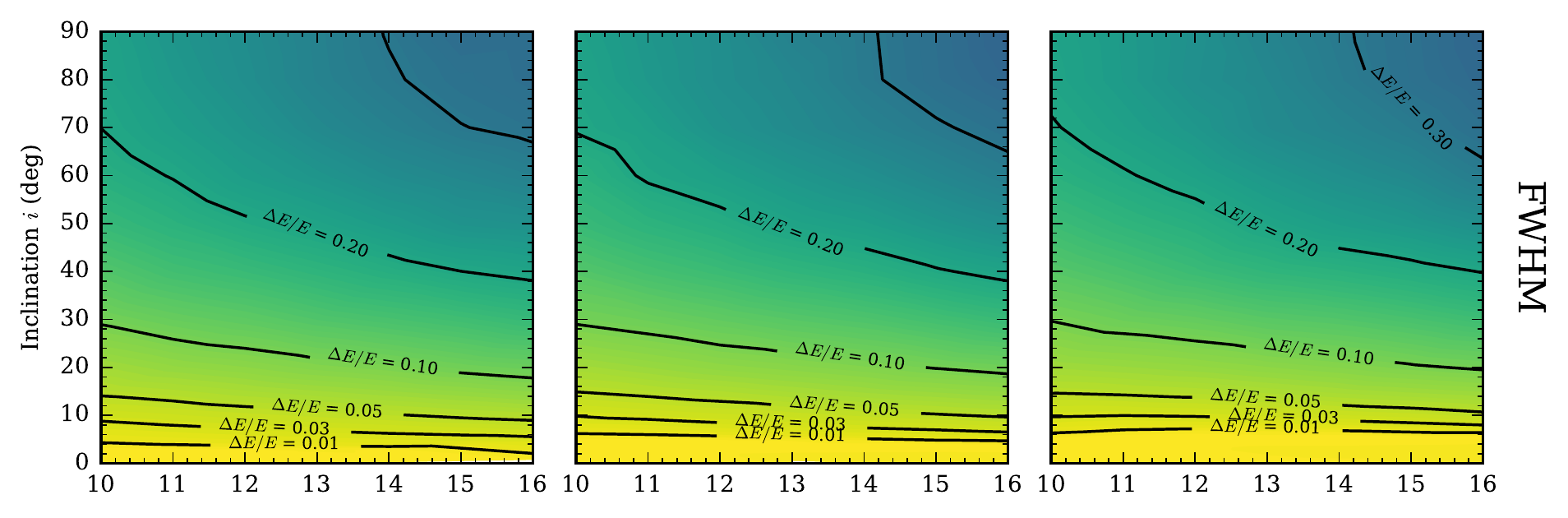}
    \includegraphics[width=18cm]{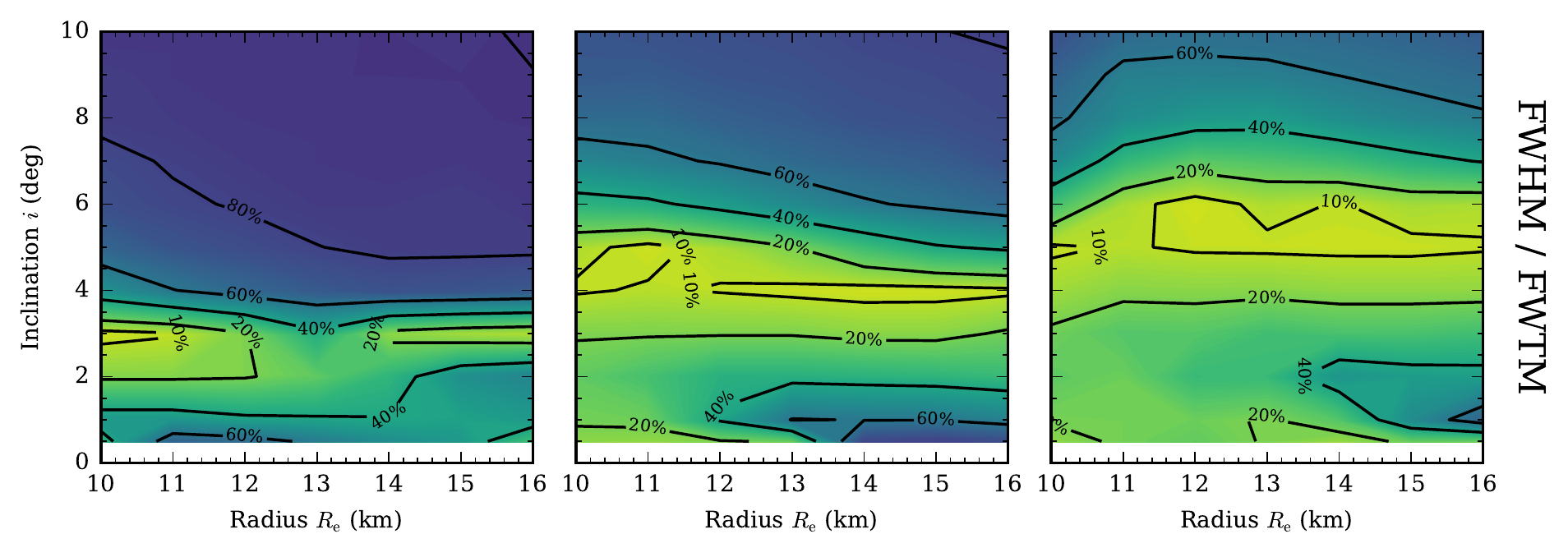}
\caption{\label{fig:fwhm}
    Line profile full width at tenth-maximum (top row) and full
width at half-maximum (middle row) as a function of radius and observer inclination computed for different neutron star configurations of $M=1.1\,\Msun$, $1.5\,\Msun$, and $1.8\,\Msun$ spinning at $600\,\mathrm{Hz}$.
    Additionally, the bottom row shows the ratio of these two,
which can be used to quantify the width of the spiky part of the line profile.
    Note the different inclination scale on the bottom row that is used to highlight the region $i < 10\deg$, where the profile evolves from a smooth to a spiky shape.
}
\end{figure*}

Next, we study the energy-dependence of the stellar flux by computing the observed energy distribution $F(E)$ of photons emitted from the stellar surface at a single energy $E'$ as measured in the comoving frame.
In order to minimize any source of error, we use \textsc{arcmancer} in this section to solve the geodesics in all of the subsequent calculations.
Because of the variation of the redshift across the surface of the star that is caused by Doppler boosting and because of the oblate shape of the star, the observer sees a range of energies \citep{OP03,BML06,CMB06}.
Following \citealt{Baubock15}, we are interested in constraining the convolution (smearing) kernel $\mathcal{G}(E,E')$ defined via
\be
F(E) = \int I'(E') \mathcal{G}(E,E') dE',
\ee
where we have dropped the time and angle dependency of the specific intensity $I'$ and have explicitly written all quantities to be functions of the energy.
It follows from Eq. \eqref{eq:fluxint} that the actual flux of photons with an observed energy $E$ and emitted energy $E'$ is then
\begin{align}\begin{split}
    F(E) &= \iint \frac{ I'(E') }{ (1+z)^3 } \frac{bdb d\chi}{D^2} \\
         &= \iiint \frac{I'(E') }{(1+z)^4} ~ \delta \left( E - \frac{E'}{1+z} \right)  \frac{bdb d\chi}{D^2} dE',
\end{split}\end{align}
where $\delta(x)$ is the Dirac delta function.
The convolution kernel, or the so-called line profile, we are after is then 
\be
\mathcal{G}(E,E') =  \iint \frac{1}{(1+z)^4} ~\delta \left[E - \frac{E'}{1+z} \right]  \frac{bdb d\chi}{D^2}.
\ee

Examples of line profiles are shown in Fig.~\ref{fig:line_profiles} for different space-times and star configurations.
The flux is normalized so that the emitted bolometric flux is unity, that is, the area encapsulated by the profile is one.
In each case, the star is taken to have $\nu = 700\,\mathrm{Hz}$, $R_{\mathrm{e}} = 10\,\mathrm{km}$, and $M=1.4\,\Msun$.
The observer inclination is $i=20^{\circ}$ and emission is taken to be isotropic, for simplicity.
This figure shows line profiles for spherical and oblate stars, assuming for simplicity that the exterior space-time is the Schwarzschild space-time, as well as results for oblate stars in the appropriate second-order exterior space-time, 
that is, a space-time with the appropriate quadrupole moment given by relations \eqref{eq:quad}-\eqref{eq:I}.

The line profiles computed using the \sch metric with a spherical star appear to be smooth and asymmetrical with an enhancement toward higher energies caused by the relativistic Doppler boosting \citep[see, e.g.,][]{OP03}.
For an oblate star, the increased redshift of the regions near the pole shifts the peak toward lower energies.
The resulting line profile is fairly symmetric \citep[see, e.g.,][]{BPO13}.
However, when a physically more realistic metric with a non-zero quadrupole moment is used, the high-energy part of the line profile is further enhanced.
This again results in an asymmetrical line profile.
When the value of the quadrupole moment is increased to unphysically high levels, the line profile develops a narrow peak in the high-energy part.
This effect highly depends on the observer inclination relative to the rotation axis of the star, however.

We now study the line profile shape in full detail using the \textsc{bender} code.
In order to fully map the change in the line profile shape as a function of observer inclination, we calculated different profiles for three different cases: $M=1.1\,\Msun$, $1.5\,\Msun$, and $1.8\,\Msun$.
Here we consider only rapidly spinning stars and hence set the spin to $600\,\mathrm{Hz}$, which is close to the maximum value observed for AMPs.
For each mass, the equatorial radius $R_{\mathrm{e}}$ and observer inclination $i$ were taken to span the full range from $10$ to $16\,\mathrm{km}$, and $0$ to $90\deg$, respectively.
Examples of the observed line profiles are shown in Fig.~\ref{fig:sweep} for $i=5\deg$, $10\deg$, $20\deg$, $40\deg$, $60\deg$, and $90~\deg$.
From here it is easy to see that the profile appears to be smooth at almost all observer viewing angles.
Only at $i \lesssim 5\deg$ , a sharp spike starts to form.
In this case, however, the actual observed width of the profile is already below $0.03 \times E$, whereas the spike is as narrow as $0.01 \times E$.
For a spectral energy feature at around $10\,\mathrm{keV}$, therefore,
a resolution of $0.1\,\mathrm{keV}$ would be needed to resolve it.

We can also try and quantify the observed effect more thoroughly by introducing the full width at half-maximum (FWHM) of the profile (i.e., the width of the profile at $F_{\mathrm{max}}/2$).
In addition, we consider the full width at tenth-maximum (FWTM) that reflects the total width of the profile (i.e., the width of the profile at $F_{\mathrm{max}}/10$).
These values are shown for different radii and observer inclinations in Fig.~\ref{fig:fwhm}.
They are also a useful measure of how the rotation would smear the observed spectra:
the FWTM gives a quantitative estimate of how widely smeared any narrow feature, such as a line or an absorption edge, would be observed.
The FWHM, on the other hand, quantifies the energy resolution needed to resolve the exact effects from rotation itself.
Finally, we can also use their ratio to describe the shape of the line profile:
the narrower (and hence localized) the line profile feature, the smaller this fraction.
For a narrow peak we expect an FWHM/FWTM of around $\sim 10\,\%$.
This ratio is shown in the bottom panels of Fig.~\ref{fig:fwhm}.
For a star rotating at $600\,\mathrm{Hz}$, a narrow line feature is visible only for observers with inclinations in a very narrow range, that is, $3\deg  < i < 6\deg$, regardless of the mass or radius of the star.

These results can be compared to the results reported in \citet{BPO13}. 
Here the line profiles using the \sch exterior metric are seen to match our calculations well.
For the profiles computed using a metric that includes corrections up to second order in $\Ob$, we see a clear deviation.
Most notably, the line profiles we compute only contain narrow features in a very restricted range of observer inclinations $3\deg  < i < 6\deg$.
However, \citet{BPO13} found narrow spectral features with observer inclinations $i \lesssim 30\deg$ with similar neutron star parameters.
A possible reason for this discrepancy can be traced back to how the value of the quadrupole moment is computed.
The value of our quadrupole moment is derived from the scaling relations, whereas \citet{BPO13} set their value of $q$ by hand.
They used the Hartle-Thorne metric \citep{HT68} in their calculations, where the quadrupole moment is given by $\qinv = -j^2 (1 + \eta)$, where $j$ is the dimensionless spin parameter (see eq. \eqref{eq:wbar}; $a$ in the notation of \citealt{BPO13}), and $\eta$ is the strength of the deviation from a spherical potential ($\eta = 0$ reduces to Kerr-like space-time).
\citet{BPO13} selected $\eta=3.3,$ which is a typical value given by the FPS EoS \citep{FPS} for a star with $M\approx1.4\,\Msun$ \citep[see][]{LP99}.
For the angular momentum they adopted $j = 0.357$, again as given by the FPS EoS at $\nu \approx 700\,\mathrm{Hz}$.
With this value, their quadrupole moment is then $\qinv \approx -0.548$.
The radius they imposed was $R = 10\,\mathrm{km}$.

However, we note that by selecting an individual EoS and setting the star's mass and angular momentum, the radius of the star is already determined for physically realistic parameter combinations.
In their case, the FPS EoS would yield a considerably different radius of $R_{\mathrm{e}} \approx 11.8\,\mathrm{km}$ \citep{CST94, LP99}.
Moreover, \citet{BPO13} did not include a contribution from the pressure quadrupole moment $\beta$ in their coordinate-invariant quadrupole expression \citep{PA12}.
On the other hand, we obtain for $M=1.4\,\Msun$, $R_{\mathrm{e}}=10\,\mathrm{km}$, and $\nu = 700\,\mathrm{Hz}$ the following values: 
$j \approx 0.275$, $q \approx -0.268$, and $\beta \approx 0.010,$ which then give $\qinv \approx -0.255$.
Hence, the value of the $\qinv$ used in \citet{BPO13} is approximately twice that of a physically realistic neutron star with $R_{\mathrm{e}} = 10\,\mathrm{km}$.

In general, the quadrupole moment is larger for a stiffer EoS, because a stiffer EoS produces a larger star and the quadrupole moment scales with the square of the radius.
For us, this scaling is taken into account by relations \eqref{eq:quad} and \eqref{eq:beta}, which are obtained by fitting a large library of EoSs \citep[see][]{BBP13, aGM14} to yield a consistent quadrupole moment at any given mass, radius, and angular velocity.  
This scaling also hides the difference between the non-rotating and rotating radii because it is formulated using the equatorial radius $R_{\mathrm{e}}$.
Alternatively,  the corresponding $R_0$ of a non-rotating configuration might be considered, for which $R_0 \le R_{\mathrm{e}}$ for any given $\Ob$.
This distinction between rotating and non-rotating radii is important as EoS modeling for the cold dense matter inside neutron stars is typically done assuming non-rotating radii.
In this particular comparison, our $j$ and $\qinv$ are therefore smaller because we require that the radius be $10\,\mathrm{km}$.
The line profile emerging from a such a star is shown with a red solid line in Fig.~\ref{fig:line_profiles} and is not seen to develop a narrow core. 
For an oblate star, we need to artificially increase $q$ by a factor of $4$, so that $q = -1.07$, in order for the line profile to produce a narrow core, as seen in the red dashed line in Fig.~\ref{fig:line_profiles}.
In conclusion, we are only able to reproduce the narrow peak with a large observer inclination of $20\deg$, shown in fig. 2 of \citet{BPO13}, by artificially increasing the value of $q$. 

\citet{BBP13} subsequently revised their calculations and recomputed their observed line profile in Hartle-Thorne metric with values of the quadrupole moment $q$ originating from a similar physically consistent empirical parameterization.
In this case, \citet{BBP13} still found a narrow spectral feature in the line profile for a neutron star with $R_{\mathrm{e}} = 10\,\mathrm{km}$, $M=1.4\,\Msun$, and $\nu = 700\,\mathrm{Hz}$, similar to the parameters used in \citet{BPO13} (see fig. 5 \citealt{BBP13}).
However, the observer inclination was not specified.
Based on the results we present in Figs. \ref{fig:sweep} and \ref{fig:fwhm}, however we can say that the formation of a narrow peak for these neutron star parameters is only possible in a very limited range of observer inclinations, of around $i \sim 5\deg$.

\subsection{AMP pulse profiles}\label{sect:AMPs}

From here on, we move to time-dependent ray tracing problems by considering pulse profiles from AMPs.
Here a hot spot on the stellar surface is emitting, and the star is seen to rotate with a frequency of $\Omega$.
The internal accuracy of the calculations is only set by the error tolerance of the numerical integration of the flux.
Hereafter we use a relative tolerance of $5 \times 10^{-3}$.
The results here were obtained using both the split Hamilton-Jacobi propagator and \textsc{arcmancer,} and they match given the numerical error tolerance.
For simplicity, we only show the results obtained by the split Hamilton-Jacobi method in the following discussion.

The definition of the differential surface element is given by Eq. \eqref{eq:dS_subs} and hence correctly includes the $\lgamma$ factor.
This comparison is crucial in order to verify that all of the physics is correctly incorporated in the formulations of the given methods.
Results between the two methods are therefore expected to agree up to the numerical tolerance.

First, a general comparison of the ray tracing algorithm with the S+D approximation was made using the \sch metric.
For simplicity, only spherical stars were considered here.
The main parameters were the stellar mass $M = 1.6\,\Msun$, the stellar radius $R = 12\,\mathrm{km}$, the observer inclination $i = 60^{\circ}$ , and the colatitude of the spot $\theta_{\mathrm{s}} = 50^{\circ}$.  
The effective radiation temperature was set to $T_{\mathrm{eff}} = 2\,\mathrm{keV}$.  
The distance to the source was assumed to be $D = 10\,\mathrm{kpc}$.  
We defined a circular spot with an angular radius $\rho$ of either $1$ or $30$ degrees.
Here the spot size is defined using its angular size in the corotating coordinate system.
The angular distribution of the radiation corresponds either to an isotropic blackbody with constant intensity or to an atmosphere
dominated by electron scattering, that is, the Hopf profile.

The light curves are computed in 128 time bins.  Zero time $t = 0$ corresponds to the moment when the spot center crosses the plane defined by the spin axis and the direction to the observer.  
We computed curves for the five following quantities: monochromatic photon flux ($\mathrm{ph}\,\mathrm{cm}^{-2}\,\mathrm{s}^{-1}\,\mathrm{keV}^{-1}$) at the observer energies $E = 2,~6,$ and $12\,\mathrm{keV}$, bolometric photon flux ($\mathrm{ph}\,\mathrm{cm}^{-2}\,\mathrm{s}^{-1}$), and the bolometric energy flux ($\mathrm{erg}\,\mathrm{cm}^{-2}\,\mathrm{s}^{-1}$).

\begin{figure*}
\centering
\includegraphics[width=18cm]{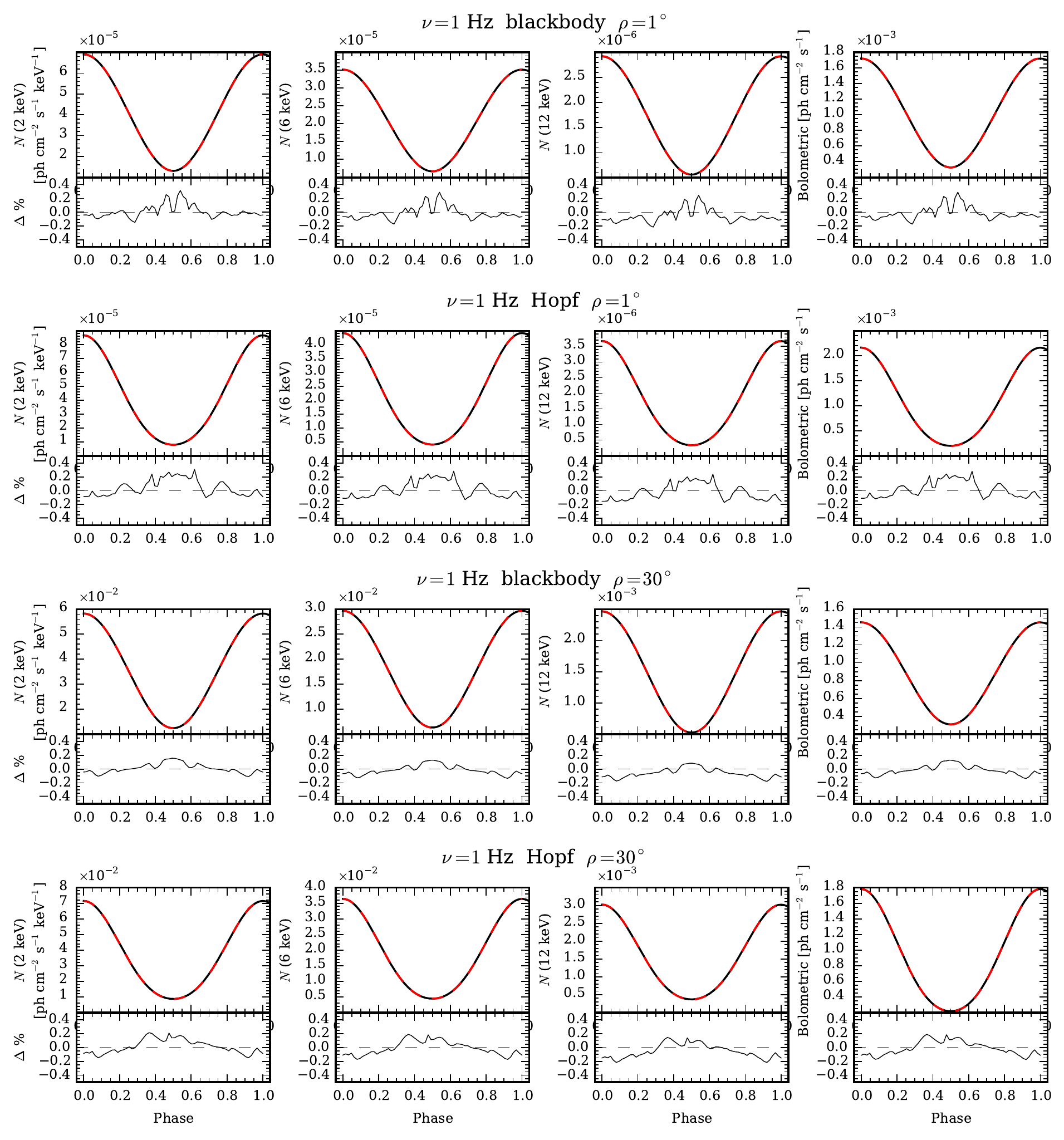}
\caption{\label{fig:sch_comp1}
  Light-curve comparisons for \sch space-time with a slowly rotating spherical star ($R = 12\,\mathrm{km}$, $M = 1.6\,\Msun$, $\nu = 1\,\mathrm{Hz}$, $i = 60^{\circ}$, $\theta_{\mathrm{s}} = 50^{\circ}$, $\rho = 1^{\circ}$, and $T_{\mathrm{eff}} = 2\,\mathrm{keV}$) emitting according to a blackbody or Hopf profile with a spot size of either $1$ or $30$ degrees.
    The black solid line shows the pulse profiles computed using the S+D approximation (forward-in-time method; see text), and
the red dashed line is a profile computed with the code presented here.
  The lower panel shows the residuals as $\Delta = (\mathrm{model_{S+D}}/\mathrm{model} -1) \times 100\%$.
}
\end{figure*}

\begin{figure*}
\centering
\includegraphics[width=18cm]{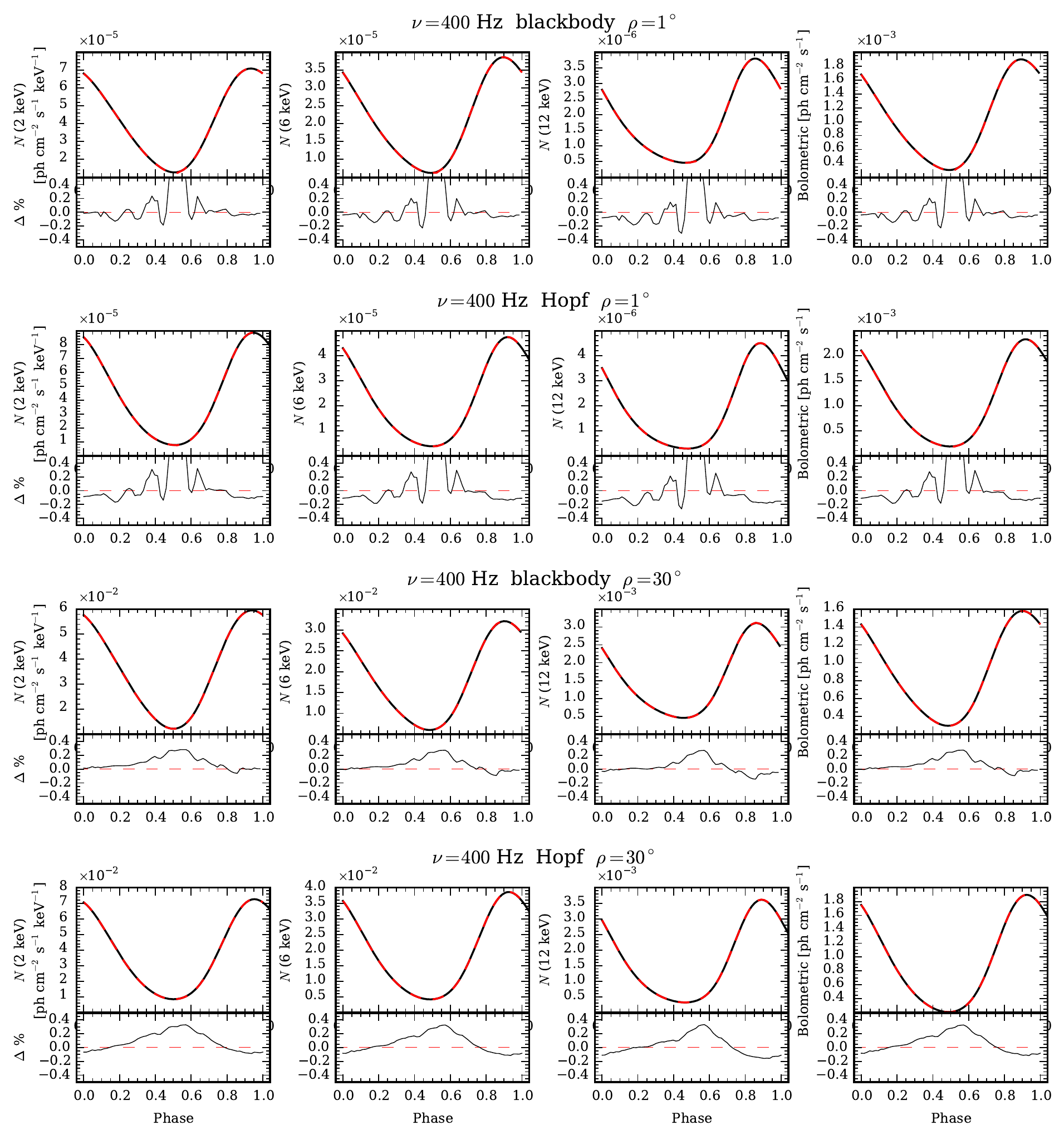}
\caption{\label{fig:sch_comp400}
  Light-curve comparisons for \sch space-time with a rapidly rotating spherical star ($\nu = 400\,\mathrm{Hz}$).
  The other parameters and symbols are the same as in Fig.~\ref{fig:sch_comp1}.
}
\end{figure*}

The comparison of these light curves is shown in Fig.~\ref{fig:sch_comp1} for a slowly rotating star ($1$ Hz) and in Fig.~\ref{fig:sch_comp400} for a fast-rotating star ($400$ Hz).
In practice, comparing the profiles for slow rotation only tests our ray tracing routines because special relativistic effects (Doppler boosting, angle aberration, and so on) are negligible.
The overall agreement of the two different methods is excellent, and from here, a baseline accuracy of about $<0.2\%$ relative error is obtained for the mapping of quantities between image plane and stellar surface.
No large deviation between isotropic and Hopf profile is detected either, indicating a good agreement in our emission angle computations and formulation.
Similarly, when rotation is increased and special relativistic effects start to play a role, we are usually able to reproduce the pulse profiles down to $<0.3\%$ relative error, except near $\phi_{\mathrm{e}} \sim 0$.
Here the tilt of the spot increases, and even though the absolute error remains the same, the relative error grows because the observed flux is increasingly lower for a more inclined spot.
This situation is numerically expensive when integrating the observed flux from the NS image.
In this case, we set set a bound on the number of flux integrand evaluations (typically $\sim 10^7$ function calls) that in effect set an absolute error for the flux.
It is then only in these rare cases that our integrator does not respect the relative error criteria set by us.

\begin{figure*}
\centering
\includegraphics[width=18cm]{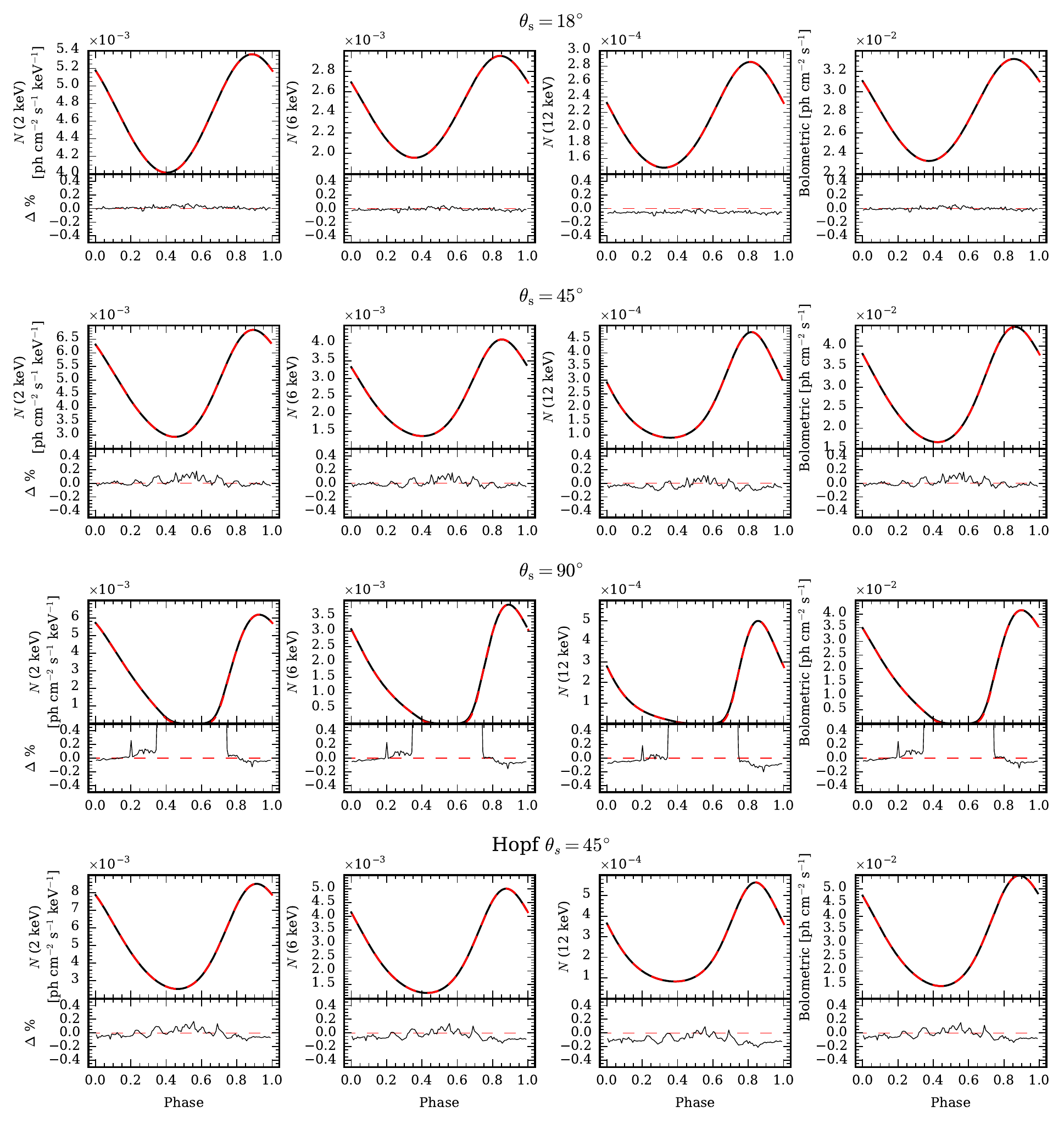}
\caption{\label{fig:osch_comp700}
  Light-curve comparisons for oblate \sch space-times with three different spot colatitudes: $\theta_{\mathrm{s}} = 18^{\circ}$, $45^{\circ}$ , and $90^{\circ}$.
  Additionally, the bottom row shows the comparison of the pulse profile for an electron-scattering atmosphere for $\theta_{\mathrm{s}} = 45^{\circ}$.
  The parameters of the star, the spot, and the observer are $R_{\mathrm{e}} = 12.0\,\mathrm{km}$, $M = 1.4\,\Msun$, $\nu = 700\,\mathrm{Hz}$, $i = 45^{\circ}$ , and $\rho = 10^{\circ}$.
  The other parameters and symbols are the same as in Fig.~\ref{fig:sch_comp1}.
  }
\end{figure*}

Next we compare emission from oblate stars.
The surface here is defined using the radius function \eqref{eq:radf}, but the star is still embedded in a symmetric \sch space-time.
The parameters we used are an equatorial radius $R_{\mathrm{e}} = 12\,\mathrm{km}$ (in the usual \sch metric), a neutron star mass of $M = 1.4\,\Msun$, an extreme rotational frequency $\nu = 700\,\mathrm{Hz}$, an observer inclination $i=45^{\circ}$, and a spot angular size of $\rho = 10^{\circ}$.
The effective temperature of the radiation was again taken to be $T_{\mathrm{eff}} = 2\,\mathrm{keV}$, and the distance to be $D = 10\,\mathrm{kpc}$.
Here the spot size is defined in a corotating spherical coordinate system on top of a unit sphere and is then projected onto the oblate inclined surface.
To trace the effects of the changing surface, we considered the spot in three different locations at colatitudes of $\theta_{\mathrm{s}} = 18^{\circ}$, $45^{\circ}$, and $90^{\circ}$.
Additionally, we considered a spot with angle-dependent emission intensity.
This was done using the electron-scattering atmosphere with $\theta_{\mathrm{s}} = 45^{\circ}$.

A comparison of the oblate light curves is shown in Fig.~\ref{fig:osch_comp700}.
Again we obtain an excellent agreement with the forward-in-time method, with similar errors as in the spherical case (relative error $< 0.2\%$).
This agreement is of course expected since our method is general and does not depend on the shape of the emitting surface.
The only large deviation is again seen when the spot is viewed from an extreme angle for $\theta_{\mathrm{s}} = 90^{\circ}$, just before the occultation.
We therefore conclude that the two methods, forward-in-time and backward-in-time, agree all the way up to the numerical tolerance.

\begin{figure*}
\includegraphics[width=18cm, height=20cm]{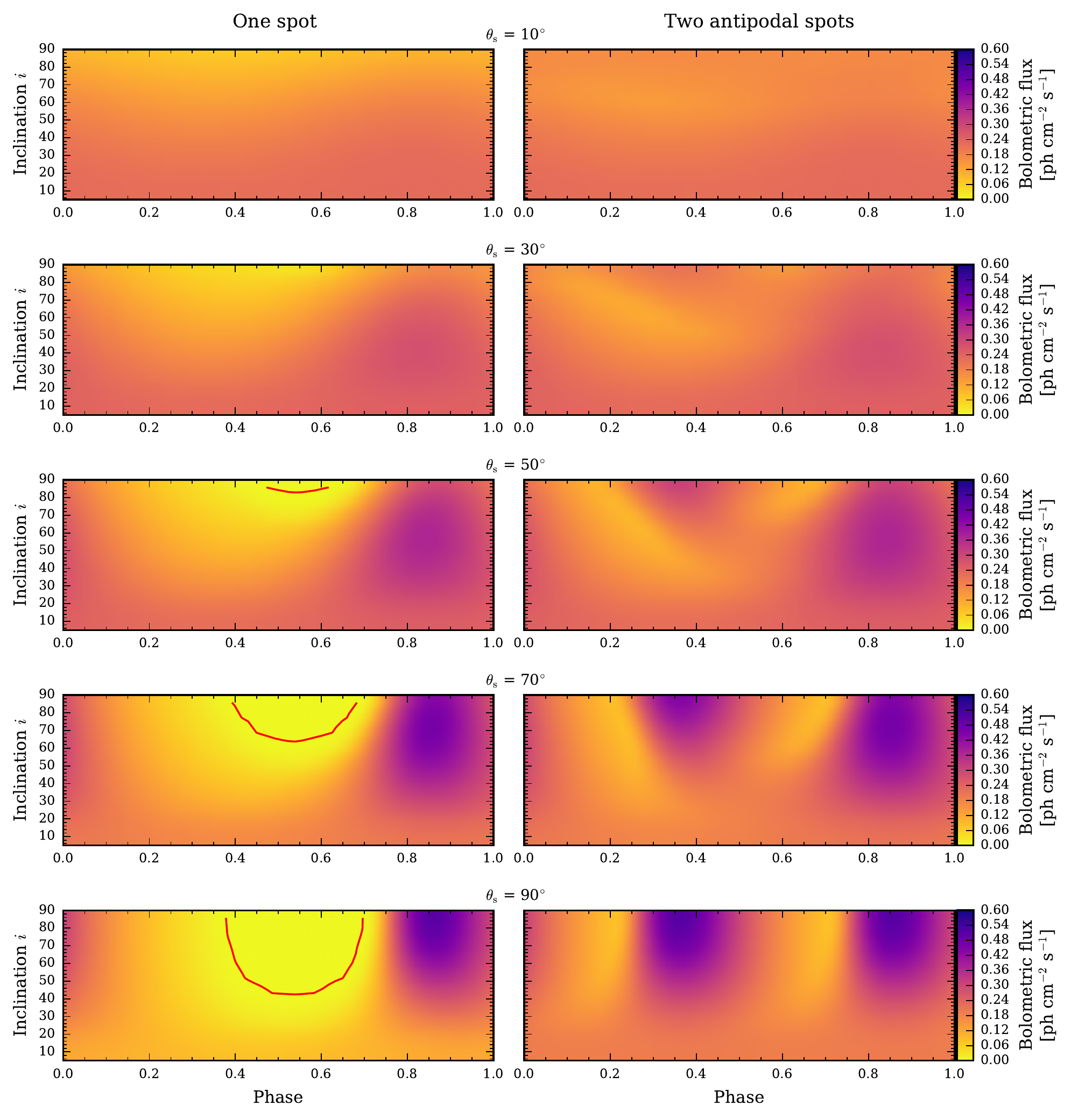}
\caption{\label{fig:skymap}
    Skymaps of the emitted radiation as produced by a rapidly rotating oblate AMP with one or two antipodal spots.
    The star is taken to have $R_{\mathrm{e}} = 15\,\mathrm{km}$, $M=1.6\,\Msun$, and $\nu = 600\,\mathrm{Hz}$.
    The spot sizes are $\rho = 10^{\circ}$ , and the emission is coming from an isotropic blackbody with $T_{\mathrm{eff}} = 2\,\mathrm{keV}$.
    The red curves enclose the area where occultation is observed.
  }
\end{figure*}

\begin{figure}
\includegraphics[width=8.5cm]{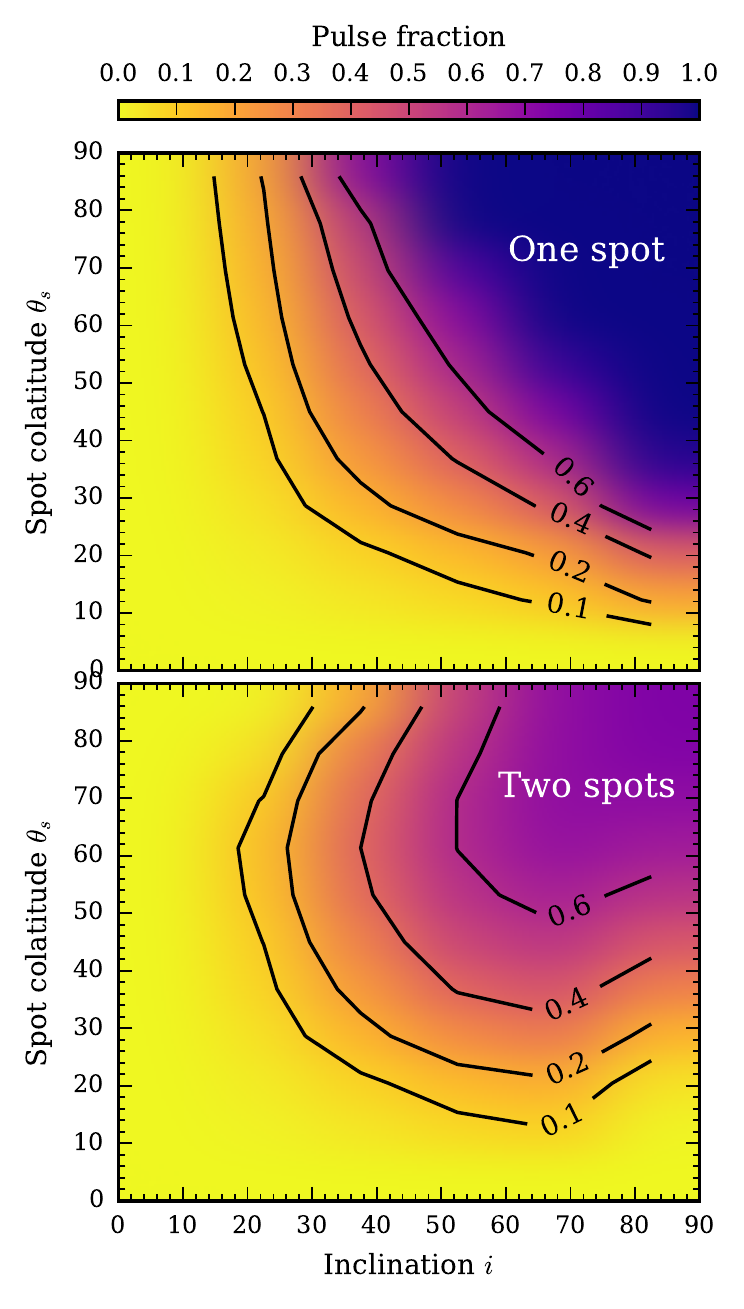}
\caption{\label{fig:pulsefracs}
    Pulse fractions observed from a rapidly rotating oblate AMP.
    The parameters of the star and the spot(s) are the same as in Fig.~\ref{fig:skymap}.
  }
\end{figure}

After the comparisons, we calculate as a last example full skymaps of the emerging radiation from rapidly rotating oblate AMPs, as shown in Fig.~\ref{fig:skymap}.
The emission from the AMP is shown for all possible observers and is mapped to the vertical axis using the observer's inclination angle $i$.
The horizontal axis of the map is the usual pulse phase.
The brightness of the skymap is proportional to the received bolometric photon number flux.
Taking a slice of the skymap at one particular value of $i$ produces the light curve as seen by the observer at that inclination.
The calculations here were made for an extreme case of an NS with $R_{\mathrm{e}} = 15\,\mathrm{km}$, $M=1.6\,\Msun$, and $\nu = 600\,\mathrm{Hz}$.
We considered a spot size of $\rho = 10^{\circ}$ , with varying colatitudes ranging from near the pole at $\theta_{\mathrm{s}} = 10^{\circ}$ to the equator at $\theta_{\mathrm{s}} = 90^{\circ}$.
We considered the cases of one spot and two antipodal spots with isotropic beaming and blackbody emission with $T_{\mathrm{eff}} = 2\,\mathrm{keV}$.
As a result, we can see that full occultations are only observed with one spot.
From here it is also easy to see the variation in phase of the flux maxima and minima when the viewing angle of the observer changes.
The effect becomes most prominent with two spots located at $\theta_{\mathrm{s}} \sim 50-70^{\circ}$ (second antipodal spot at $\theta_{\mathrm{s,2}} \sim 110-130^{\circ}$), and the minimum is seen to change from around phase of 0.2 all the way to 0.4.

We also show the corresponding strength of the observed pulsations for each inclination and spot colatitude combinations in Fig.~\ref{fig:pulsefracs}.
Here the color of the image corresponds to the pulse fraction computed by
\be
f_p = \frac{F_{\mathrm{max}} - F_{\mathrm{min}}}{F_{\mathrm{max}} + F_{\mathrm{min}}},
\ee
where $F_{\mathrm{min}}$ and $F_{\mathrm{max}}$ are the minimum and maximum values in the bolometric light curves, respectively.
From here the symmetry between $\theta_{\mathrm{s}}$ and $i$ becomes obvious as the lines of constant amplitude appear almost symmetric against switching between $x$ and $y$ axis.

\section{Summary}\label{sect:summary}
We have presented a detailed study of radiation emerging from and near rotating compact objects.
A framework of formulae for solving this problem was derived in a fully general relativistic manner. 
The formulae were then specialized to the context of rotating neutron stars.

First, we gave a detailed description of the second order in rotation space-time metric in Sect. 2.1.
The space-time we used has a non-zero coordinate-invariant mass quadrupole moment $\qinv$. 
The components $q$ and $\beta$ of $\qinv$ are defined via approximate relations for a wide span of neutron star masses, radii, and spins, following \citet{aGM14}.
When the rotation increases, the star also starts to deviate from a sphere because the gravitational force weakens on the equator because the centrifugal force increases.
An approximate relation for the resulting oblate spheroidal shape of the star was again obtained \citep{MLC07, aGM14} and was implemented in Section \ref{sect:oblate} for an easy but consistent description of the surface.

Second, we derived a new approximate ray tracing approach using the so-called split Hamilton-Jacobi method (also known as super-Hamiltonian
method).
This derivation was presented and discussed in detail in Sections \ref{sect:hamjac} and \ref{sect:raytracing}.
Instead of using the geodesic equation that is a second-order differential equation, we separated the Hamilton-Jacobi equation using a third constant of motion known as Carter's constant.
The method is exact up to first order in rotation (Kerr-like space-time with frame-dragging effects) but remains sufficiently accurate also for second order in rotation because deviations caused by the quadrupole moments are small.
Formulating the components of the four-momentum vector like this has the useful feature that the polarization of the radiation can easily be taken into account \citep[see, e.g.,][]{cha, dexter2016}

Third, we gave a thorough description of the calculations related to the actual emission of the radiation.
Effects such as redshift, Doppler boosting, and emission angle of the photon were discussed in a fully general relativistic manner in Sections \ref{sect:redshift_angle} and \ref{sect:emission}. 
In the special relativistic formulation \citep[see, e.g.,][]{PB06}, the calculations were made in a flat space-time and were then Lorentz-boosted to the rotating relativistic frame.
In Section \ref{sect:coords} we presented a derivation of the solid-angle element that we defined using a rotating coordinate system.
The purpose of this was to clarify some common misunderstandings in the literature of how this transformation from the corotating to the static coordinate system can be achieved.
We also briefly discussed in Sect. 2.8 the actual intensity of the emerging radiation and presented an iterative method to solve the Chandrasekhar-Ambartsumian integral, along a new approximate polynomial expansion that is related to the angle-dependent electron scattering atmosphere.
We then described the actual intensity of the emerging radiation and used as a simple model the angle-dependent electron-scattering atmosphere presented in Section \ref{sect:angular_distr}.
Numerical methods for solving all of the presented equations were then laid out in Section \ref{sect:num_methods}.

Finally, in Section \ref{sect:appl} we presented some applications of the framework.
A connection to previous work in the literature was also made, when possible.
As a first simple example, we showed how the image of an NS is formed in curved space-time.
Next, we studied the energy-dependent emission by considering the emerging line profiles.
Most notably, we concluded that when a consistent formulation is used to describe how the increasing eccentricity of the star is coupled to the related quadrupole moments, the resulting line profiles develop a narrow core only at an observer inclination of $i \sim 5\deg$.
Otherwise, the smearing kernels are smooth functions.
The effect of the rotational smearing on the observed energy spectra can be estimated using the FWHM and FWTM of the kernels,
which we computed for all observer viewing angles.
We then studied AMP pulse profiles extensively and thoroughly, and we compared our results to results obtained using existing special relativistic methods found in the literature.
Here the agreement between the two methods was found to be excellent when the correct differential surface area element presented in Sect. \ref{sect:coords} was used.
Last, we computed full skymaps of the radiation that emerges from rapidly rotating AMPs, taking into account the oblate shape of the star and the quadrupole moments of the space-time metric.

\section*{Acknowledgments}

\small{
We sincerely thank various people that helped us to improve the formulation of the theory and the actual paper by engaging the authors in helpful discussions: 
Most notably we would like to thank Juri Poutanen, Fred Lamb, Jean in 't Zand, Cole Miller, and Sharon Morsink for their help and comments.
We also offer special thanks to Tuomo Salmi for producing the pulse profiles for the code comparison.
JN acknowledges support from the University of Turku Graduate School in Physical and Chemical Sciences.
PP acknowledges support from the Academy of Finland, grant no.~1274931.

This work benefited from discussions at the International Symposium on Neutron Stars in the Multi-Messenger Era in May 2016, supported by the National Science Foundation under Grant No. PHY-1430152 (JINA Center for the Evolution of the Elements).
}

\bibliographystyle{aa}

\end{document}